%% file: CLIP_Qscore_Slonimczyk_Karapsin_v0.tex
\documentclass[a4paper,11pt]{article}

\usepackage{url}
\usepackage{graphicx}
\usepackage{natbib}
\bibpunct{(}{)}{,}{a}{,}{,}
\usepackage{amsmath}
\usepackage{amsfonts}

\usepackage[labelfont=bf]{caption}
\usepackage[labelfont=bf]{subcaption}

\usepackage[english]{babel}
\usepackage[latin1,utf8]{inputenc}
\usepackage{listings}
\usepackage{tikz}

\usepackage{commath}
\usepackage{ctable}
\usepackage{color,colortbl,array}
\definecolor{Gray}{gray}{0.9}
\usepackage{afterpage}
\usepackage[a4paper,margin=1in]{geometry}
\usepackage{tabularx}
\usepackage[most]{tcolorbox}
\usepackage{xcolor}
\usepackage{pifont}

\newtcolorbox{systembox}{
    colback=blue!3, colframe=blue!40!black, title=System Prompt,
    fonttitle=\bfseries\small, fontupper=\ttfamily\scriptsize, arc=1mm
}

\newtcolorbox{humanbox}{
    colback=blue!3, colframe=blue!40!black, title=User Query,
    fonttitle=\bfseries\small, fontupper=\ttfamily\scriptsize, arc=1mm
}
\newcommand{\chkmrk}{\textcolor{green!60!black}{\checkmark}}
\newcommand{\xmrk}{\textcolor{red}{\ding{55}}}

\author{Fabi\'an Slonimczyk\thanks{International College of Economics and Finance, HSE. \texttt{fslonimczyk@hse.ru}.} \and Danila Karapsin\thanks{Yandex}}
\title{Measuring Product Quality Using Images: The CLIP Q-Score and an Application to Real Estate}
\date{August 2026}

\begin{document}
\bibliographystyle{kbib}

\maketitle

\begin{abstract}
The CLIP Q-score is a novel, safe, fully reproducible, and computationally efficient method for extracting objective product quality metrics from visual data using Contrastive Language-Image Pre-training. We introduce the technique and provide an extensive application to real estate data from an online platform ($\sim500,000$ images). Our open-source metric aligns with LLM assessments and proves to be a powerful predictor of housing market prices for both sales and rentals. We also show that a higher CLIP Q-store is associated with better liquidity (reduced time on the market), especially for properties on sale.
\end{abstract}

\begin{flushleft}
{\footnotesize JEL classification: C45; C81; R31; D83\newline
  }
\end{flushleft}

\setlength{\parskip}{0.1cm}
\newcommand{\cov}{\mathrm{cov}}
\newcommand{\1}{\mathbf{1}}
\newcommand{\E}{\mathbf{E}}
\newcommand{\D}{\mathbf{\Delta}}

\newpage
\begin{quote}
    \flushright
    \footnotesize Seeing is believing
\end{quote}

\begin{quote}
    \flushright
    \footnotesize One look is worth a thousand words
\end{quote}

%\begin{quote}
%    \flushright
%    {\footnotesize \songti 百闻不如一见  (Bǎi wén bù rú yī jiàn)}  \\
%    {\footnotesize [Seeing for oneself is a hundred times better than hearing from others]}
%\end{quote}

\begin{quote}
    \flushright
    {\footnotesize Un bon croquis vaut mieux qu'un long discours } \\
    {\footnotesize [A good sketch is worth more than a long speech]}
\end{quote}

\section{Introduction}

As the quotes above suggest, a single image can often convey facts, complex ideas or emotions more effectively than a lengthy verbal description. Image-based communication is pervasive even in highly textual domains: catalogues, menus, advertising, and posters are ubiquitous features of modern life.

For most consumers, being able to see how a product looks provides valuable additional information. E-commerce and other digital platforms have responded by placing increasing emphasis on image and video content.

In this paper, we develop a new technique to extract an image-based measure of product quality. Our method leverages a multimodal neural network trained using contrastive language-image pre-training (CLIP). We refer to the resulting measure as the CLIP Q-score. We provide a general introduction to the technique and then offer a step-by-step guide of using the model to obtain a quality score.

We illustrate our approach with an application to real estate valuation. Specifically, we show that our method can measure the ``luxuriousness'' of housing units based on pictures posted by owners and real estate agents on an online platform.\footnote{Alternatively, one could interpret our results as measuring the state of disrepair of the properties. Our measure is constructed such that higher scores reflect better quality.} For this purpose, we collected and analyzed data from a large online platform, including almost half a million images.

As a validation exercise, we compare the resulting CLIP Q-measure to a score obtained by querying a multimodal large language model. We show the two approaches yield scores that are highly positively correlated.  We also exploit the richness of our data and several institutional and historical facts to show that the CLIP Q-score follows reasonable patterns. First, it is highly correlated with users' self-reported ``state of repair'' of the property. Second, the U-shaped pattern of real estate quality with respect to year of construction coincides with the historical record. Third, the geographic distribution of scores aligns with anecdotal evidence regarding Moscow neighborhoods: the best real estate is in the city center and the north-western and western districts, whereas properties in the far-eastern and far-southern districts are, on average, in a worse state of disrepair. Finally, we show that properties in buildings marked for demolition under a state-sponsored program receive markedly lower scores.

Most importantly, our CLIP Q-score is a robust predictor of market outcomes. Better-looking properties are listed for significantly higher prices, even after controlling for a large number of quantifiable characteristics. We also show that, among similarly priced units, those with higher image scores are easier to rent or sell.

Our paper is related to a large and fast-growing literature that uses image-based information in economics and finance. \citet{Obaid_2020} show that an image-based sentiment score can predict market movements. \citet{Ash_et_al_2021} use computer vision to reveal systemic thematic stereotyping of minorities in the media. \citet{Gorin_Heblich_Zylberberg_2025} construct a long-run sentiment score from art works spanning centuries. \citet{voth_yanagizawa_2024} obtain measures of style from graduation pictures and document significant cultural changes. \citet{Xiaojiang_2021} use Google Street View to measure the fraction of green vegetation in street-level images, which is known to be associated with mental and physical health benefits. \citet{Jean_et_al_2016} employ satellite images to estimate poverty levels. \citet{Dzyabura_et_al_2023} use product images to predict individual item return rates.

Within real estate economics, our paper is related to recent work demonstrating that image-derived features improve automated valuation models and hedonic regressions \citep{Kostic_2020, Deng2025, Tapia2025}. It is also an evolution of past efforts trying to enrich the hedonic model with non-conventional data \citep[see, for example][]{Nowak_Smith_2017}.

Our paper is unique in that it provides a new general-purpose technique for extracting quality information from images. CLIP Q-scoring offers several advantages relative to measures derived from querying an AI model. First, the CLIP score is completely reproducible, meaning that other researchers will be able to obtain \emph{exactly} the same scores.\footnote{All the data, including the images and code used in this study, are open source.} Second, applying our method is practically free and requires minimal computing resources. Moreover, because CLIP can be run locally, it is computationally efficient. Finally, our technique does not require exporting data to inference providers, which ensures data safety and confidentiality.

Our paper also makes important contributions to the large literature studying how urban real estate markets price heterogenous property characteristics. Our study is unique in that we provide evidence for both sales and rental markets, and in that we study outcomes other than prices.

The paper is organized as follows. The next section provides a detailed explanation of the CLIP model and how we use it to derive the Q-score. In Section 3, we describe the real estate data used in our application. Section 4 validates the resulting CLIP Q-scores by comparing them to an alternative obtained by querying a multimodal LLM and presents a thorough descriptive analysis showing that the CLIP Q-score follows reasonable patterns. In Section 5, we embed the CLIP Q-score in a hedonic model predicting property prices based on their characteristics and show that our score is among the most highly predictive features. Section 6 extends the analysis to time on the market. Finally, Section 7 concludes.

\section{Methods: a CLIP-based measure of quality}

We propose a novel way to extract quality information from product images. Specifically, we take advantage of a contrastive language-image pre-trained (CLIP) model \citep{Radford_et_al_2021}.

\begin{figure}
    \centering
    \includegraphics[scale=0.25]{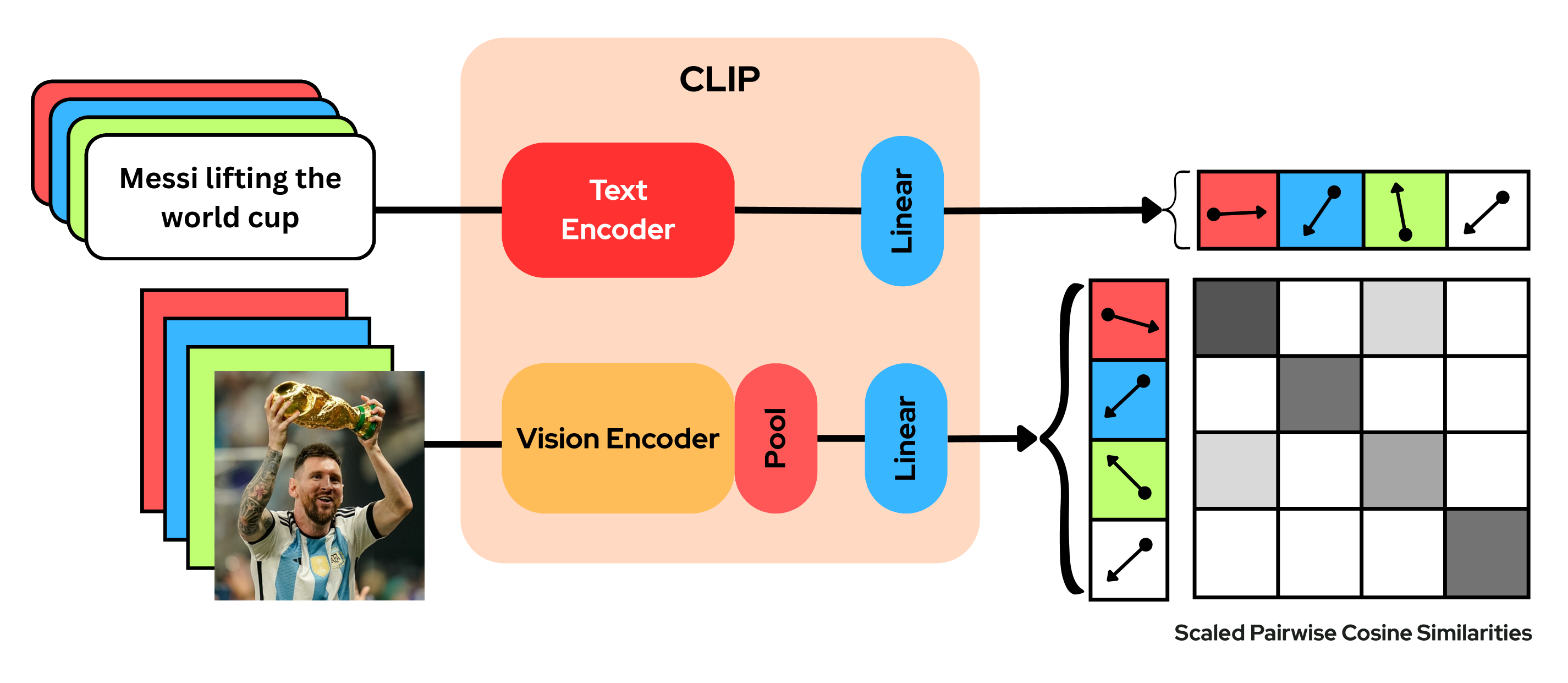}
    \vspace{-0.1in}
    \caption{CLIP Diagram}\label{fig_CLIP_diagram}
\end{figure}

CLIP is a general purpose technique to train multimodal neural networks, i.e. deep models which can accept \textbf{both} images and text as inputs.\footnote{OpenAI introduced the CLIP model to learn joint visual-textual representations \citep{Radford_et_al_2021}. \citet{Hessel_et_al_2021_clipscore} later formalized the model's outputs into ``CLIPScore''—--a reference-free metric that can be used for robust automatic evaluation of image captioning.} Figure~\ref{fig_CLIP_diagram} presents a schematic representation of the technique.

In a nutshell, CLIP combines two models: one capable of encoding text and another capable of encoding images. While many implementations are possible in principle, most current versions combine a regular transformer encoder \`a la BERT and a vision transformer (ViT). By encoding, here we mean that the models are capable of generating contextualized embeddings of their input sequences. These two embeddings ---text and image--- are then projected onto spaces of equal dimensionality by a fully-connected linear layer.

What is truly remarkable about CLIP are the ideas behind how the two models are trained. \citet{Radford_et_al_2021} realized that plentiful image data from the internet often is accompanied by a caption or another easy-to-identify short textual description.\footnote{Establishing the link between the image and the text is often facilitated by HTML tags.} In other words, it was relatively easy for their research team to create a vast training set of image-caption pairs.

With this in mind, they proposed a loss function that would reward the models when they produced similar embeddings for image-text pairs that matched, but close to uncorrelated embeddings for texts and images that were unrelated. Specifically, for each batch of training data, CLIP computes all pairwise similarity metrics between image and text embeddings. The CLIP loss function then maximizes the values in the diagonal of the similarity matrix, while minimizing the absolute value of the off-diagonal elements.

While the diagram in the figure has batch size of only four, the actual CLIP implementation works with batch sizes of 32K image-text pairs or larger. The first case in the diagram batch is the text-image pair of Messi lifting the 2022 FIFA world cup. The other cases in the batch are color-coded. Text and image elements are passed through the appropriate encoder, resulting in corresponding embeddings represented in the diagram by arrow vectors.

Formally, let the $n$-size batch of normalized (i.e. unit distance) vision and text embeddings be given by:
$$\boldsymbol{E}_{img}, \boldsymbol{E}_{txt} \in \mathbb{R}^{n,d},$$%
with $d$ being the dimensionality of the embedding space. The symmetric contrastive cross-entropy loss is calculated as follows:
\begin{align*}
  \boldsymbol{S} &= \boldsymbol{E}_{img}\cdot \boldsymbol{E}_{txt}^{\prime} \\
  \boldsymbol{L} &= \boldsymbol{S} * \tau \\
  \mathrm{loss}_{img}  &= \mathrm{colXentropy}(\boldsymbol{L}, [1,\ldots,n]) \\
  \mathrm{loss}_{txt}  &= \mathrm{rowXentropy}(\boldsymbol{L}, [1,\ldots,n]) \\
  \mathrm{loss}        &= (\mathrm{loss}_{img} + \mathrm{loss}_{txt})/2
\end{align*}
where $\boldsymbol{S}$ is the $n\times n$ similarity matrix and $\tau$ is a trainable scale parameter. The CLIP objective is operationalized by minimizing the average column-wise and row-wise cross-entropies of the logits $\boldsymbol{L}$ with respect to the labels $[1,\ldots,n]$.

Training CLIP over millions of image-text pairs, \citet{Radford_et_al_2021} obtained remarkable state-of-the-art zero-shot performance on a number of benchmarked tasks.

\subsection{The CLIP Q-score}

In this paper, we show how CLIP can effectively be used to extract quality information from product images. To illustrate the approach, we develop an application to real estate valuation.

Table~\ref{tabl_CLIP_calculation} presents a summary view, which splits the calculation into five steps. Given an image of a product, the first step requires inputting the image together with three purposely composed descriptions of the product: one strongly positive, one neutral, and one strongly negative.

For example, in our application to real estate we used the following text inputs:

\begin{itemize}
  \item[Text $+$] This is a photo of a \emph{luxurious} apartment.
  \item[Text $\sim$] This is a photo of an \emph{ordinary} apartment.
  \item[Text $-$] This is a photo of a \emph{dilapidated} apartment.
\end{itemize}

In general, any set of text inputs are bound to yield reasonable results as long as there is a meaningful enough contrast among them. It should be clear from this example that the technique can be easily adapted to other settings.

% Checking the robustness of the resulting polarity measure to alternative implementations should be considered part of the method.

\begin{table}[hbt!]
  \centering
    \caption{CLIP-based Polarity Calculation}\label{tabl_CLIP_calculation}
  \begin{tabular}{lp{0.05in}p{4.5in}}
    \toprule
    Step 1 && Input each photo together with three ($+$, $\sim$, $-$) alternative text descriptions \\ \midrule
    Step 2 && Use the text and image embeddings to obtain similarity scores: \\
           && $S(j) = \boldsymbol{e}_{img}\cdot \boldsymbol{e}_{j}, \quad j\in\{+, \sim, -\}$ \\ \midrule
    Step 3 && Use model learned temperature to get logits: \\
           && $L(j) = S(j) * \tau$ \\ \midrule
    Step 4 && Transform logits into probabilities: \\
           && $P(j) = \frac{\exp(L(j))}{\sum_{h\in\{+, -, \sim\}} \exp(L(h))}, \quad j\in\{+, \sim, -\}$ \\ \midrule
    Step 5 && Obtain polarity metric: \\
           &&  $\mathrm{polarity}_{CLIP}= P(+) - P(-)$  \\
    \bottomrule
  \end{tabular}
\end{table}

The model's output from step 1 will consist of an embedding vector for the image ($\boldsymbol{e}_{img}$) and three embedding vectors for the text inputs ($\boldsymbol{e}_{+}$, $\boldsymbol{e}_{\sim}$ and $\boldsymbol{e}_{-}$). Steps 2--4 involve simple operations meant to extract the implicit probabilities that the model assigns to each of the texts.

As explained above, CLIP is trained to produce similar image-text embedding pairs when the text is a close description of the image. By matching the product ---here an apartment--- image with three divergent descriptions of its quality, we obtain the model's implicit assessment of the product's quality.

Depending on the application, researchers might find it useful to analyze each of the probabilities independently. In fact, it is not hard to image situations in which a finer gradation of quality levels could be desirable. In this case, all that would be necessary is to incorporate additional text descriptors.

In this paper, we focus on obtaining a scalar measure of quality. Thus, in step 5 we follow the sentiment analysis literature and take the polarity ($P(+) - P(-)$) score as the final output.\footnote{The neutral text affects the polarity score through the scaling of the probabilities.} Throughout the paper, we refer to the scalar polarity measure as the CLIP Q-score.\footnote{As already mentioned, \citet{Hessel_et_al_2021_clipscore} refer to the similarity metric between the text and image embeddings as a ``CLIPScore''.}

\subsection{Implementation}

The CLIP implementation we use was trained by OpenAI.\footnote{We use the open source version found in HuggingFace, called \texttt{openai/clip-vit-base-patch32}.} It uses a ViT model that tokenizes $224\times 224$ pixel images into (49) $32\times 32$ patches. Besides the trained model, OpenAI has released a pre-processor that converts image-text batches into the correct format.\footnote{For example, larger images are uniformly scaled down to the desired dimensions. The scaling is done in a way that respects the aspect ratio. The image is first uniformly scaled down so that its shorter side matches the target dimension. The pre-processor then takes a square $224\times 224$ crop directly from the absolute center of the resized image. Anything outside of that central box is thrown away, meaning some perimeter information might be lost.}

There are other versions of CLIP, both by OpenAI and other laboratories. Given the novelty of our approach, we favored OpenAI as it was the original developer and it makes for a good benchmark. The base model with 32-sized patches is relatively imprecise but very fast, which was well suited for our application.

\section{Real Estate Data}

\begin{figure}[hbt!]
  \centering
  \includegraphics[scale=0.7]{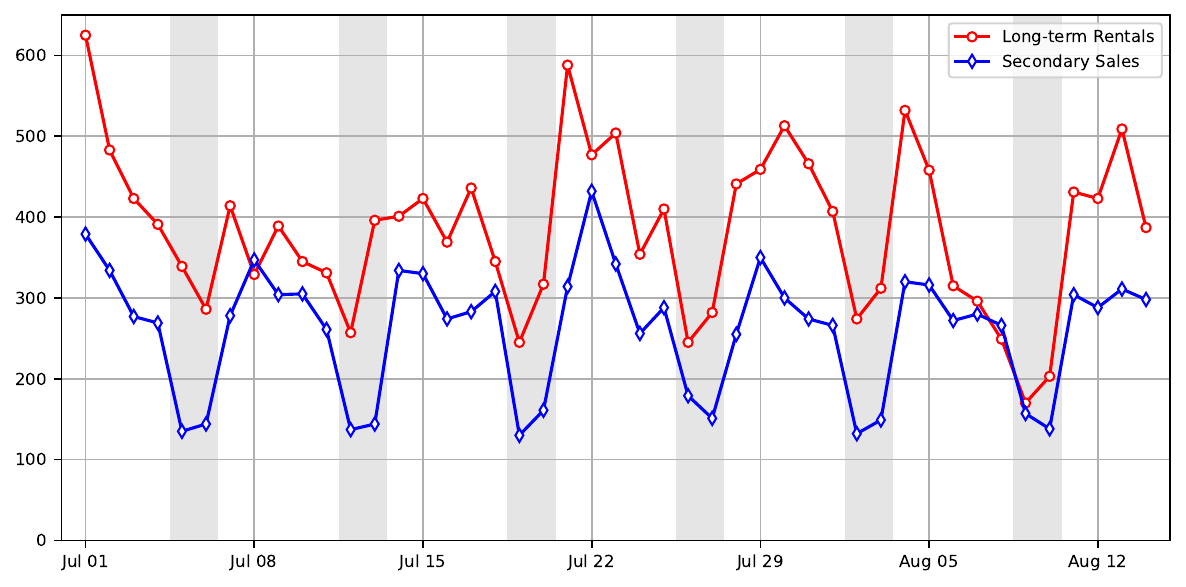}
  \vspace{-0.25in}
    {\footnotesize \flushleft \hspace{-3.0in} Notes: Weekends are shaded in gray.}
  \caption{Data Collection Results --- Post Creation Date}\label{fig_first_creation_date}
\end{figure}

The data comes from \url{cian.ru}, the main online real estate platform in Russia. We crawled the site and obtained information on every Moscow property posted for long-term rent or sale from July 1st to August 14th 2025.\footnote{The sales data analyzed here is for secondary sales only. We separately collected data for new construction but we do not use it as the image data is not informative (new construction in Russia is almost exclusively sold without any refurbishment).} Figure~\ref{fig_first_creation_date} shows the number of ad openings over the data collection period.

On average, about 400 rental and 250 sales ads were opened daily during the data collection period. There is a clear weekly pattern, however, with markedly less ads being posted during weekends.\footnote{While we know the username and contact information of the author, in general ads are posted anonymously. We do know, however, that the largest majority of ads are posted by real estate agents either individually or through major agencies' accounts. Private owners only post about 17\% of the ads.}

We followed all posts in our data set until December 5th. As a result, we have data on the complete price history for each property, as well as a precise estimate of how long the property was on the site until the post is taken down.

\begin{figure}[hbt!]
  \centering
  \includegraphics[scale=0.7]{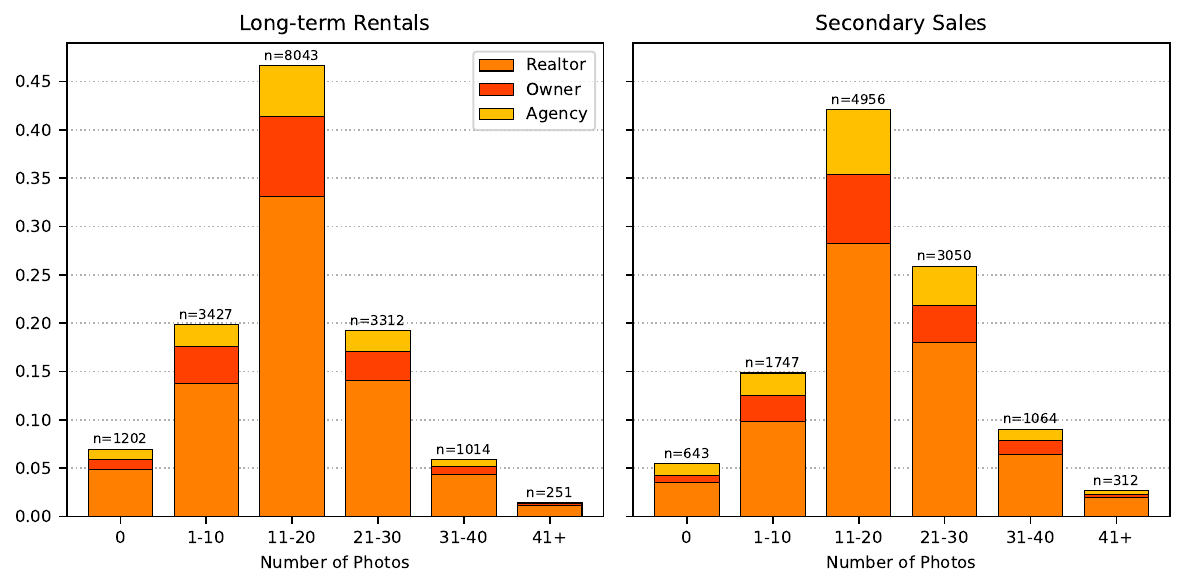}
  \vspace{-0.25in}
  {\footnotesize \flushleft \hspace{-0.05in} Notes: Moscow ads posted in \url{cian.ru} between 7/1/2025 and 8/14/2025. $N=$~17,249/11,772.}
  \caption{Images per Post}\label{fig_photos_num_barplot}
\end{figure}

\begin{figure}[hbt!]
  \centering
  \includegraphics[scale=0.7]{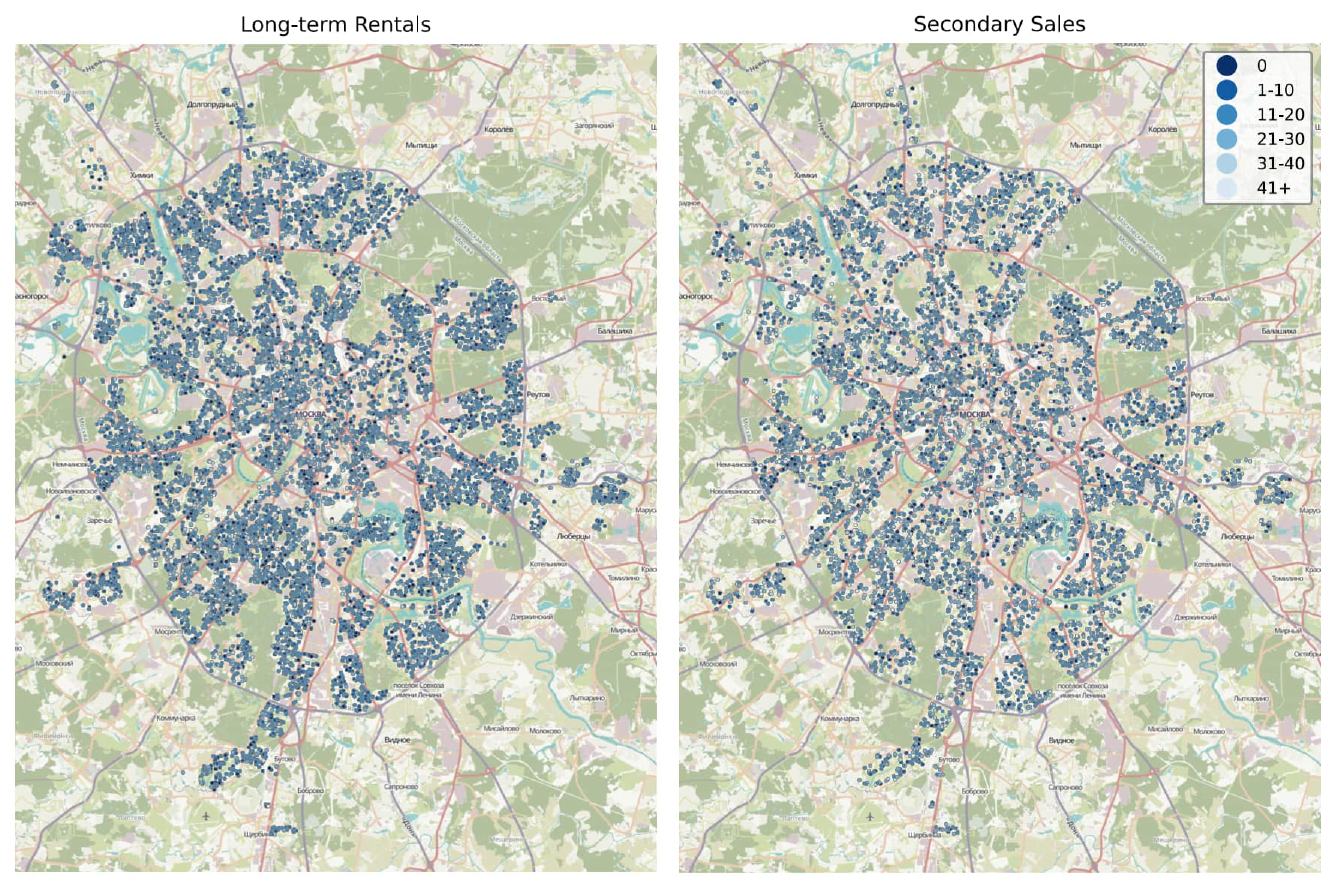}
  \caption{Property Location and Number of Images}\label{fig_photos_num_map}
\end{figure}

Most important for this study, we collected all images of the property available on the site. Figure~\ref{fig_photos_num_barplot} shows the distribution of images per post in our data. There is a small fraction (roughly 7\% for rentals and 5.5\% for sales) of posts without images. The number of images does not seem to depend on whether the author is a professional. In figure~\ref{fig_photos_num_map}, we present the geographic distribution of the properties, color-coded according to how many images the corresponding post had. Properties whose advertisements we collected are spread all throughout the city. Given the small number of image-less posts, and the fact that there is no clear pattern in their distribution, we only keep observations with images in the analysis that follows.

For each post, we collected all available property descriptors including square meterage, number of rooms and bathrooms, ceiling height, floor number, and others. We also have information on the building: year of construction, construction materials, and detailed geographic coordinates.

There are several important geographic determinants of a property's desirability. We focus on distance to the city center, distance to a metro/light-rail station, distance to a major green area/waterfront, and distance to an industrial zone.\footnote{In Moscow, unlike some other cities, the city center is considered a highly desirable area. Industrial zones are shaded in light purple in the maps of figure~\ref{fig_photos_num_map}. They include railway depots, energy and heating generation sites, and military zones. Parks are shaded in green.}

\begin{table}[htb!]
  \centering
  \caption{Sample Means}\label{tabl_descriptive_stats}
  {\tiny
    \include{tabl_descriptives}

  }
\end{table}

In table~\ref{tabl_descriptive_stats}, we present sample means for all available descriptors. With a couple of exceptions, apartments for rent and for sale are quite similar on average. Rentals tend to be slightly smaller and have less rooms and bathrooms. The main difference between the two types of posts can be seen in the reported ``state of repair'' of the property. Almost one in five apartments for sale have received no capital investment post-construction. In contrast, only a few apartments offered for long-term rental are in this category.

\subsection{Image Data}

All in all, we have almost half a millon images.\footnote{Specifically, we have 275,477 and 216,271 images for rentals and sales respectively.} All of them are in JPEG format and are almost exclusively RGB color images.\footnote{There are only 248 (rentals) and 185 (sales) grayscale images, all of them depicting architectural floor plans.}

\begin{figure}[hbt!]
  \centering
  \includegraphics[scale=0.7]{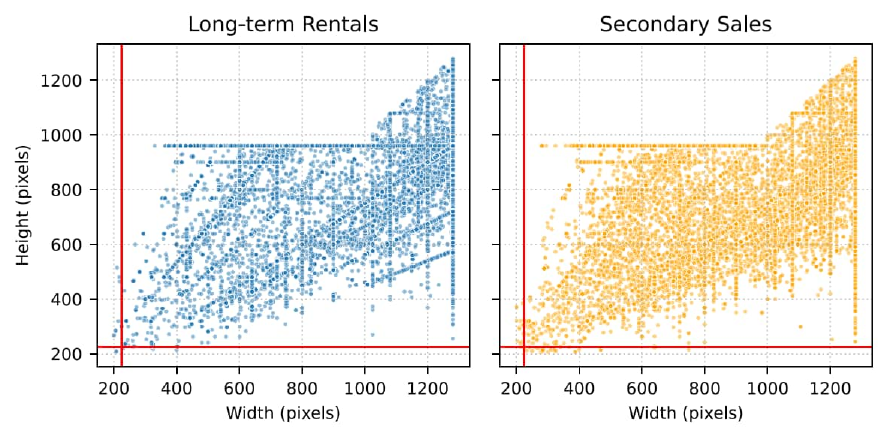}
  \vspace{-0.25in}
  {\footnotesize \flushleft \hspace{-0.4in} Notes: The red lines are set at the 224 pixel level.}
  \caption{Image Dimensions}\label{fig_photos_dimension}
\end{figure}

Figure~\ref{fig_photos_dimension} shows the dimensions of the image data. Photos vary widely both in their resolution (the total number of pixels) and in their height/width ratios. The red lines in the figure are set at the 224 pixel level, i.e. our CLIP implementation pre-processor's target.

In addition to resolution, images also vary in their quality. We collected information on three alternative quality measures. First, we obtained a measure of each image's sharpness. Specifically, we calculate sharpness via the Laplacian Variance. This method passes a Laplacian filter over the image to detect edges. A high ($1000+$) variance means sharp edges, while a low variance ($<100$) means blurry transitions.

Second, we applied a commonly used machine learning model called BRISQUE.\footnote{This stands for Blind Reference-less Image Spatial Quality Evaluator.} The BRISQUE model compares the image's pixel distribution against that of a training dataset of pristine images. The score ranges between 0 and 100, where 0 is perfect quality and 100 signifies a terribly blurry or distorted image.

Finally, we measure the image's brightness using its HSL average lightness channel.  A value of 0 means all RGB channels are flat zero (pure black). A value of 1 means all channels are max-ed out at 255 (pure white).

\begin{table}[htb!]
  \centering
  \caption{Image Quality Descriptives}\label{tabl_photo_descriptives}
  {\tiny
    \include{tabl_photo_descriptives}

  }
\end{table}

In table~\ref{tabl_photo_descriptives} we present descriptive statistics for the different quality measures. There is a relatively equal distribution of landscape ($width>height$) and portrait images. Most importantly, these statistics show that photos tend to be of high quality on average. Both the sharpness and BRISQUE scores are within the good range of values for a large majority of images. Finally, the brightness statistic is also near the sweet spot value of 50\% on average.

\begin{figure}[hbt!]
  \centering
  \includegraphics[scale=0.7]{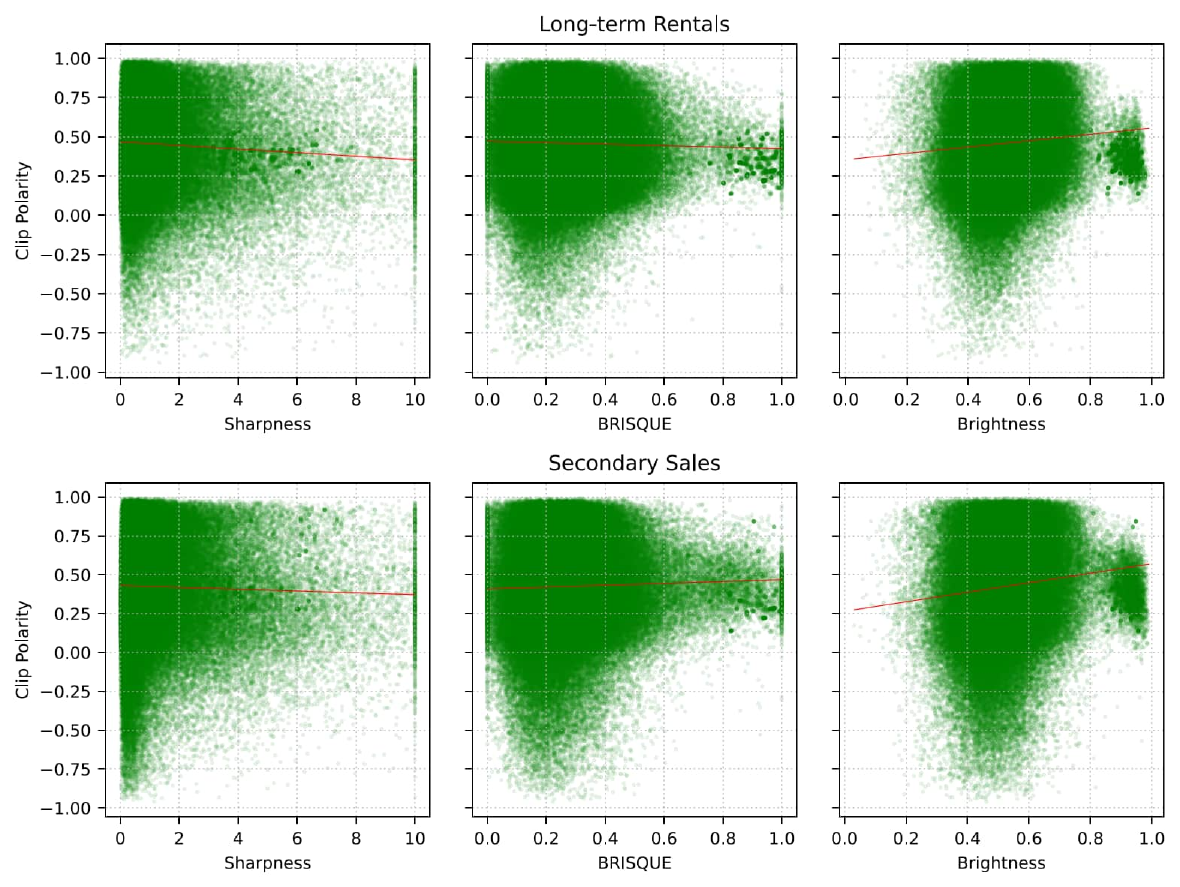}
  \vspace{-0.25in}
  {\footnotesize \flushleft \hspace{-1.2in} Notes: The red lines show the linear fit.}
  \caption{Image Quality and CLIP Polarity}\label{fig_quality_CLIP_scatters}
\end{figure}

The final line in table~\ref{tabl_photo_descriptives} presents statistics for our CLIP polarity score. While the scores are solidly positive on average, there are also plenty of negative values.

It is reasonable to wonder whether the CLIP score captures true attributes of the property or instead merely reflects photo quality. Figure~\ref{fig_quality_CLIP_scatters} shows that the latter is unlikely to be the case. While there are mild correlations among the quality measures and the CLIP score, the patterns are quite weak.

To test this proposition formally, we estimated a linear regression of the CLIP polarity score on the quality measures. Given that the CLIP pre-processor uniformly scales down images to $224\times 224$ pixels, it could also be that case that CLIP polarity depends on the aspect ratio or resolution of the image. Thus, we also include these variables in the model:

\begin{align}\label{reg_clip_quality}
  \mathrm{CLIP} &= \alpha + \beta_1 Resol + \beta_2 Ratio + \beta_3 Sharp + \beta_4 BRISQUE + \beta_5 Bright + \varepsilon
\end{align}

\begin{table}[htb!]
  \centering
  \caption{Regression Results --- CLIP Polarity and Image Quality}\label{tabl_reg_clip_quality}
  {\tiny
    \include{tabl_regression_clip_quality}

  }
\end{table}

Table~\ref{tabl_reg_clip_quality} presents estimation results for equation~\eqref{reg_clip_quality}. All the image descriptors have statistically significant partial correlations with the CLIP polarity measure. However, over 97\% of the dependent variable variance remains unexplained by the regression model. Moreover, considering the scales involved, the regression coefficients are tiny, which speaks to a practically meaningful relationship.\footnote{For example, even a full unit shift in the BRISQUE score, with all other regressors kept constant, would only marginally affect the predicted CLIP score.}

In light of these results, we conclude that photo quality and other technical aspects related to digital imaging cannot account for the wide variation in CLIP scores across images and properties. Regardless, in the analysis that follows we will employ an adjusted CLIP polarity score meant to remove the potential effects from these factors. Specifically, the adjusted score is the average residual from the regression model~\eqref{reg_clip_quality} for each property.

%  IMAGE CONTENT TBD
%
%A final important points involves the content of the images. A simple visual exploration reveals that there are three types of images: photos of the property's interior, photos of the exterior, and floor maps.\footnote{Exterior photos include window/balcony views taken from the inside, as well as street level photos of the entrance and surroundings.} Unfortunately, the image data does not come with a label or any other indicator that could be used to determine what image is what. In order to have an idea of the relative importance of the different types of images, we took a random sample of 120 ads (half rentals, half sales) and manually classified them.
%
%\begin{table}[htb!]
%  \centering
%  \caption{Image Contents}\label{tabl_photo_content_stats}
%  {\tiny
%    \include{tabl_photo_content_validation}
%  }
%\end{table}
%
%Table~\ref{tabl_photo_content_stats} shows the results from this exercise. For long-term rentals, the large majority of the images show the interior of the property; it is relatively rare to find an image of a floor plan; and only about 15\% of image show the exterior. Ads of properties for sale, on the other hand, more frequently include pictures of the view from a window or the park next door. Reasonably, it is also more likely to find floor plans in these ads.

\section{Validation}

In this section, we present a series of results meant to show that the CLIP polarity score is a good measure of quality. First, we compare the CLIP Q-scores with an alternative measure obtained from querying a multimodal LLM. We show that the scores obtained from this alternative approach are highly positively correlated with the CLIP polarity score.

Second, we present descriptive evidence showing that the CLIP Q-score shows a number of desirable patterns. Properties with `elite' observable characteristics ---higher ceilings, higher floor, designer finish, etc.--- have higher CLIP Q-scores. Properties situated in buildings scheduled for demolition have much lower scores. Properties in traditionally `posh' neighborhoods of the city have higher average scores. Finally, there is U-pattern of CLIP Q-scores and year of construction that is consistent with the known history of the Soviet period and the early transition period.

\subsection{Quality Scoring using a Multimodal LLM}

State-of-the-art LLMs offer the possibility of submitting queries that include images and other media, not just text. In other words, they are multi-modal.

The exact API for querying such a model is rapidly evolving. In appendix table~\ref{table_LLM_query}, we present an example customized for LLaMA4, a recently released multimodal LLM \citep{meta2025llama4}. We submitted all ads in batches of 20--30 and setting the temperature parameter to zero.\footnote{We experimented with different inference providers and finally settled for Novita. Processing all ads took almost two weeks and cost about \$250. An important limitation of LLaMA4 is that a maximum of 10 images can be submitted. For ads that exceeded this limit, we picked ten images selected at random. If a submission failed to generate a score, we re-submitted once. In the end we managed to obtain scores for about 90\% of ads.}

\begin{figure}
    \centering
    \includegraphics[scale=0.6]{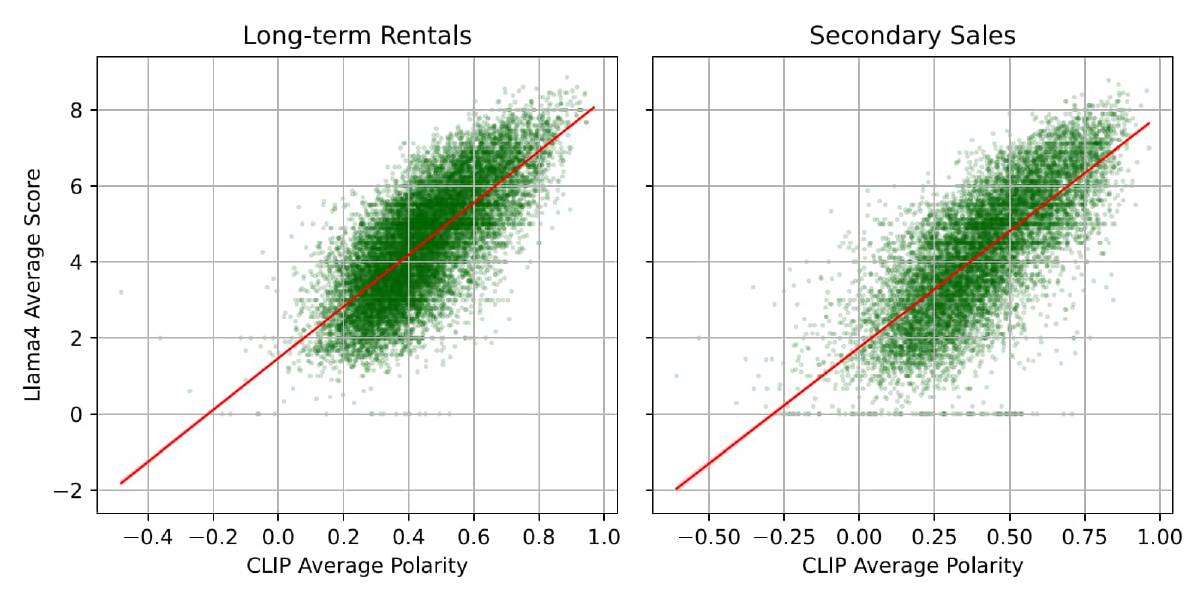}
    \vspace{-0.275in}
    {\tiny \flushleft \hspace{-1.6in} Notes: $N=$~14,129/10,226 (ads successfully scored by LLaMA4).}
      \caption{Comparing Approaches --- LLaMA4 and CLIP Polarity}\label{fig_scatter_llama4_clip}
\end{figure}

In appendix figure~\ref{fig_LLama4_CLIP_kdensities}, we present estimates of the distributions of scores for the LLM and the CLIP approaches. Despite the stark difference in approaches, the two distributions look remarkably similar. In figure~\ref{fig_scatter_llama4_clip}, we directly compare the two approaches. There is a clear positive relationship between the LLaMA4 and CLIP Q-scores.

While the LLM approach might be considered to have some advantages in terms of easiness of implementation, it also has some serious drawbacks. First, LLMs yield results that are, at best, reproducible on average. Even setting the temperature to zero, the LLM generating function is known to produce a range of outcomes for identical inputs \citep[see recent investigations by][]{Atil_et_al_2025,Tamba_2026}. In contrast, because it does not involve sampling of any sort, CLIP inference is completely reproducible.

In addition, the LLM approach requires submitting data to a service provider, which has data protection implications and makes this approach relatively expensive and slower.

\subsection{Descriptive Analysis}

Here we present some descriptive evidence that the CLIP polarity score display predictable patterns for a quality measure. First, figure~\ref{fig_clip_property_size} shows the relationship between our CLIP Q-score and property size. Larger square meterage is always reflected in a higher score, regardless of the number of rooms. In the second panel (b), we see that the pattern remains the same if conditioning on the number of rooms. Finally, the lower panel (c) shows that properties with more than one bathroom ---a relative rarity in Moscow--- also have markedly higher scores.

\begin{center}
\vbox{
 \captionsetup{type=figure} % Sets environment context to "figure"
 \caption{CLIP Q-Score and Property Size}\label{fig_clip_property_size}
    \begin{subfigure}{0.98\textwidth}
        \centering
        \includegraphics[scale=0.6]{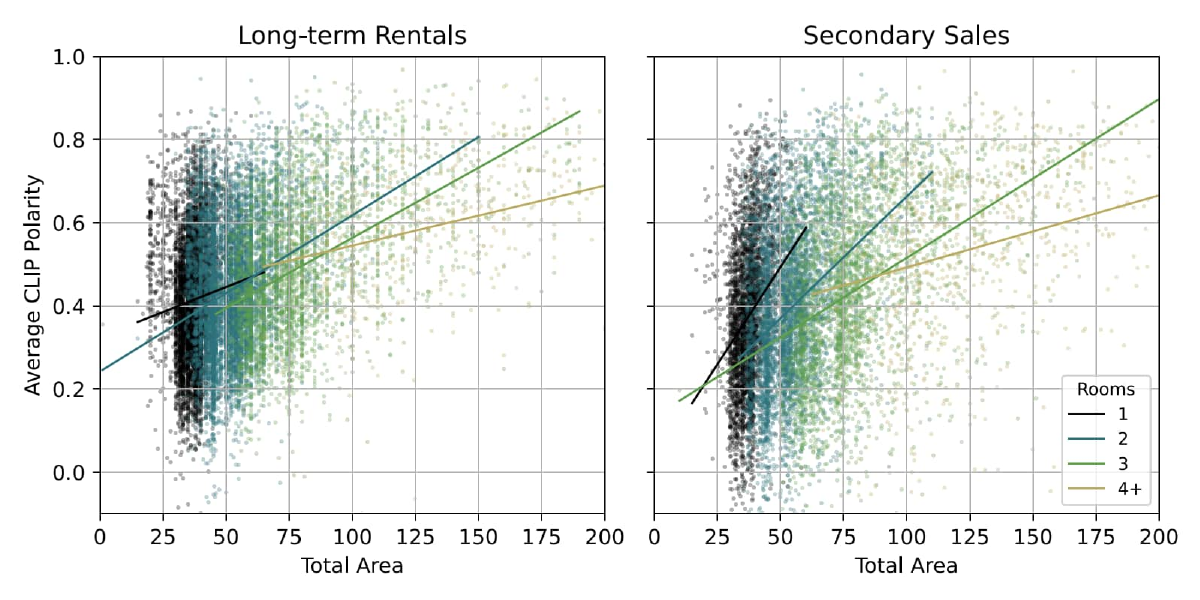}
        \caption{Total Area by Room Number}
    \end{subfigure}
    \begin{subfigure}{0.98\textwidth}
        \centering
        \includegraphics[scale=0.6]{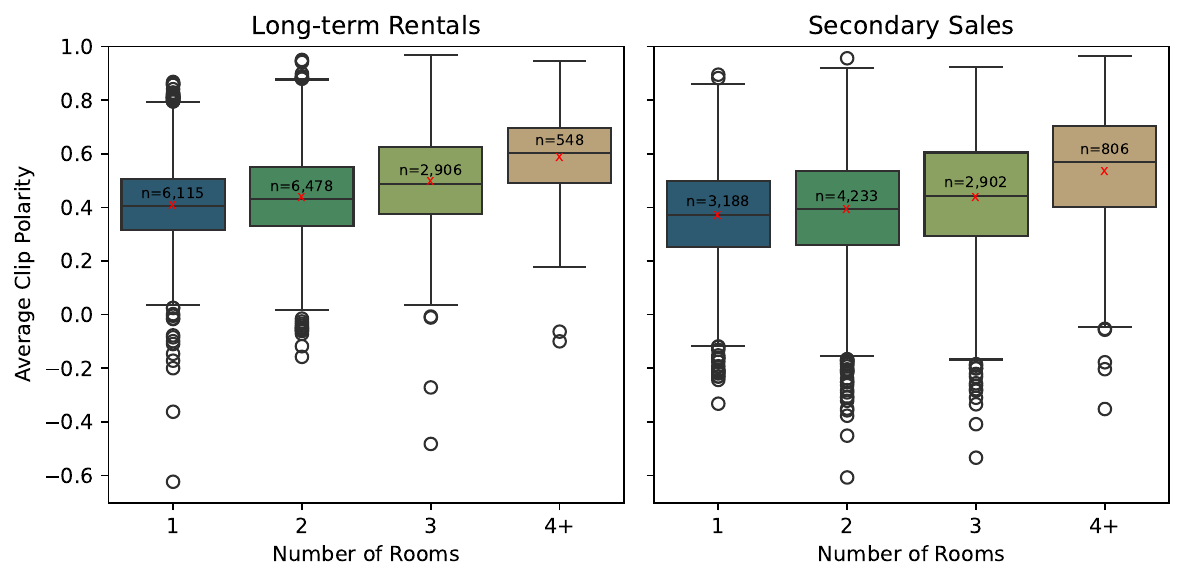}
        \caption{Number of Rooms}
    \end{subfigure}
    \begin{subfigure}{0.98\textwidth}
        \centering
        \includegraphics[scale=0.6]{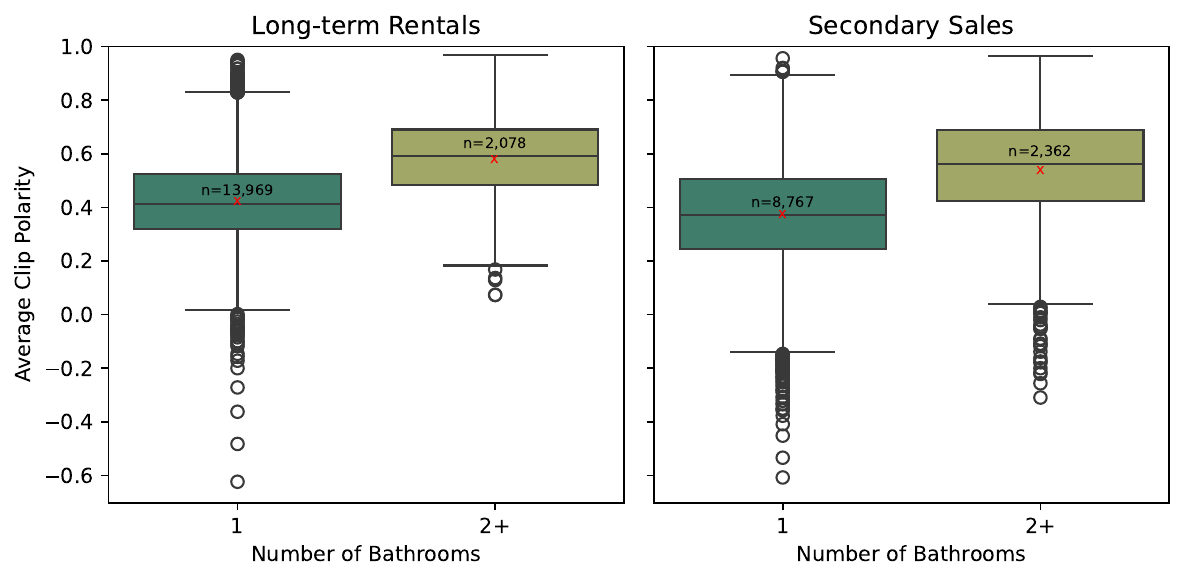}
        \caption{Number of Bathrooms}
    \end{subfigure}
}
\end{center}

Figure~\ref{fig_clip_property_chars} shows how the CLIP Q-score correlates with other proxies of quality. Apartments that are in higher floors, that have taller ceilings, and that have more sophisticated refurbishments all have better scores on average.

\begin{center}
\vbox{
    \captionsetup{type=figure}
    \caption{CLIP Q-Score and Property Characteristics}\label{fig_clip_property_chars}
    \begin{subfigure}{0.98\textwidth}
        \centering
        \includegraphics[scale=0.6]{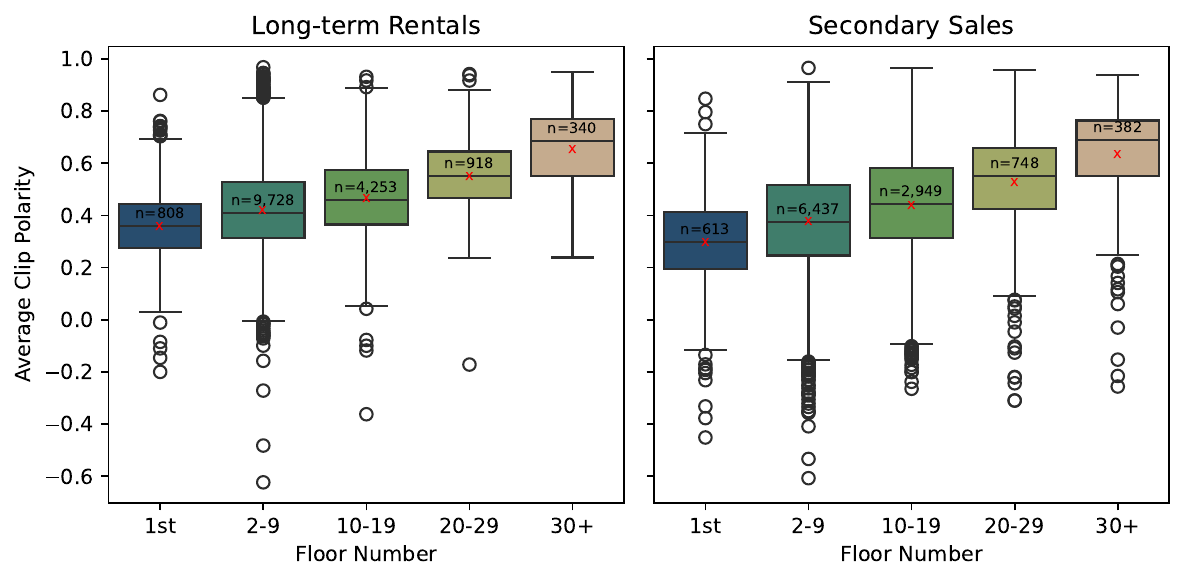}
        \caption{Floor Number}
    \end{subfigure}
    \begin{subfigure}{0.98\textwidth}
        \centering
        \includegraphics[scale=0.6]{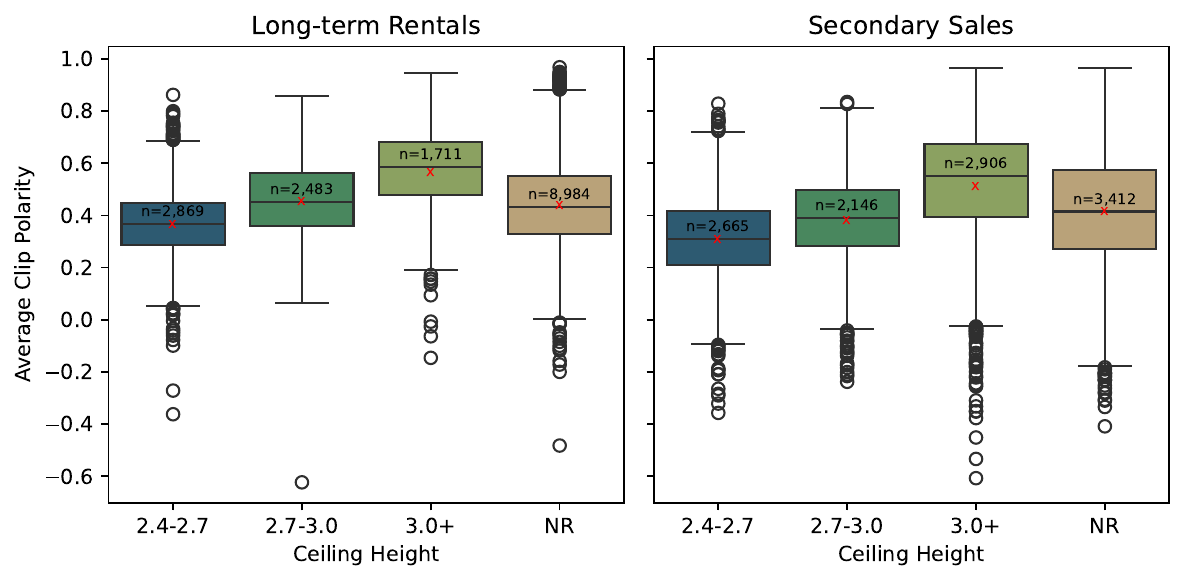}
        \caption{Ceiling Height}
    \end{subfigure}
    \begin{subfigure}{0.98\textwidth}
        \centering
        \includegraphics[scale=0.6]{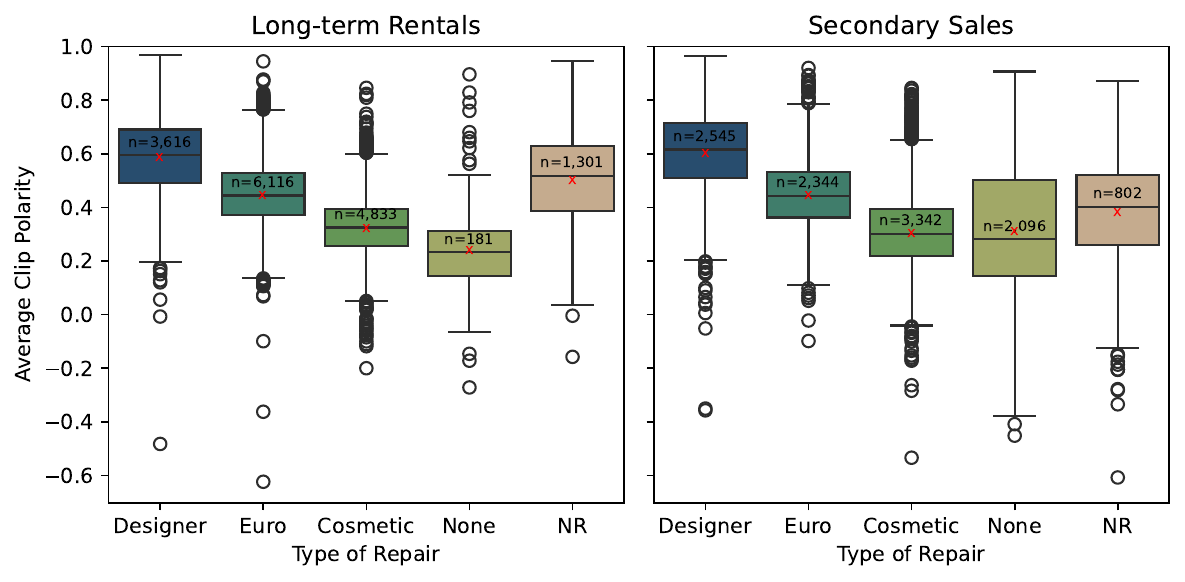}
        \caption{Repair Condition}
    \end{subfigure}
}
\end{center}

\begin{figure}
  \centering
  \includegraphics[scale=0.6]{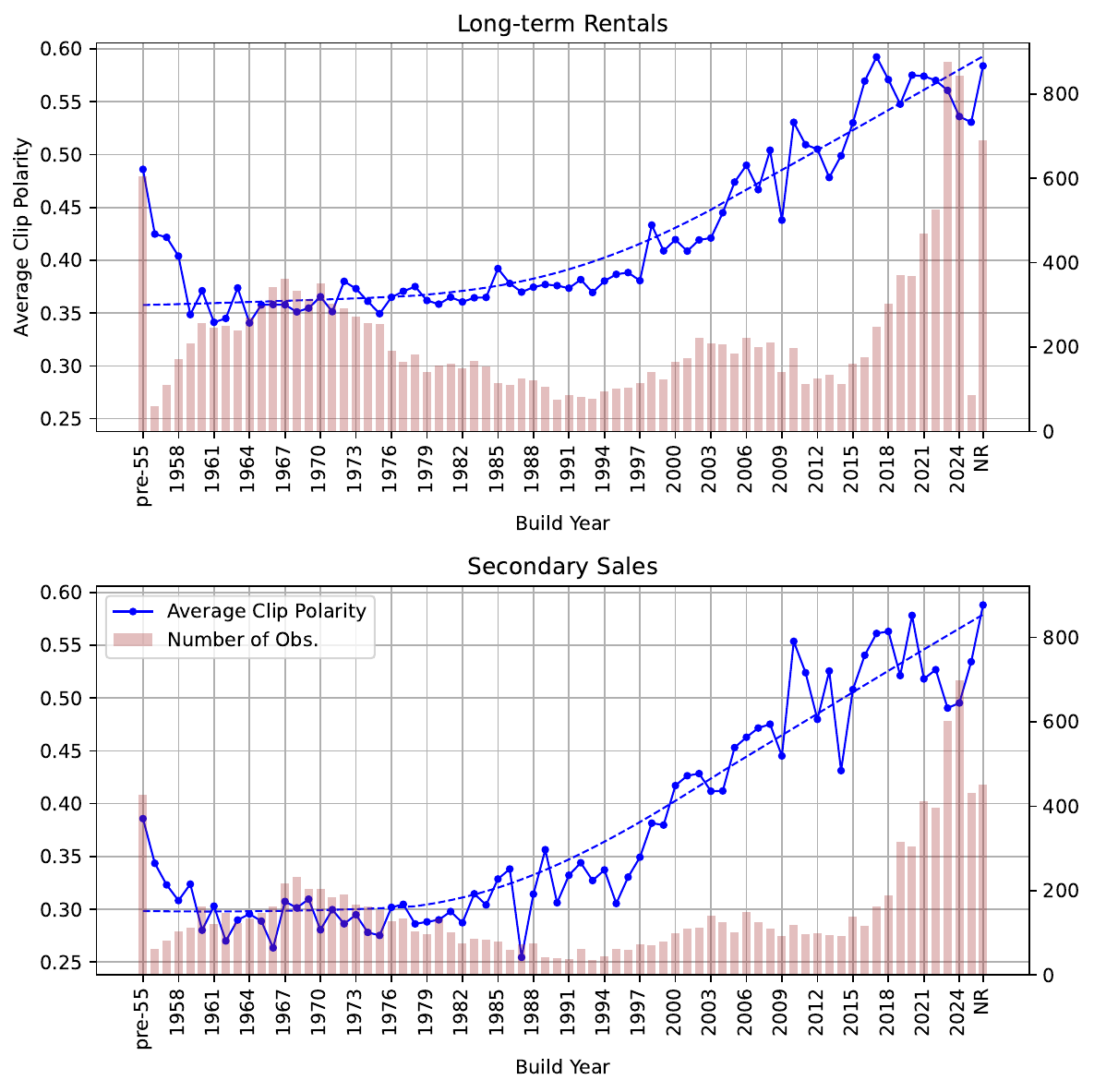}
  \caption{Year of Construction and CLIP}\label{fig_clip_year_construction}
\end{figure}

One institutional peculiarity of Moscow is that a large share of the housing stock was built during the Soviet era. Figure~\ref{fig_clip_year_construction} sorts the properties in our sample by construction year. Both properties for rent and for sale show a similar pattern. Apartments from the pre-war era that survive to this day tend to have been heavily refurbished and in good condition. Apartments built during the post-war Soviet era are in relatively worse condition. Finally, apartments built after the transition to the market economy are in the best relative condition.\footnote{The figure also shows the relative drought of new construction during the Soviet crisis period 70s--80s and the rapid increase in construction volume post-2000.}

\begin{figure}
  \centering
  \includegraphics[scale=0.6]{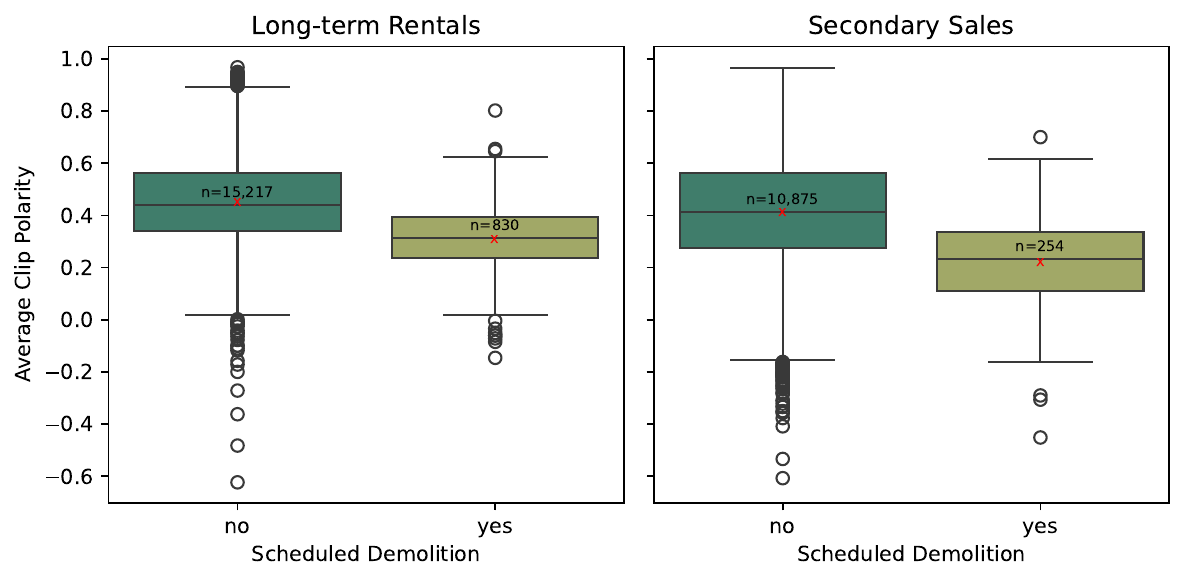}
  \caption{Moscow's Demolition Program and CLIP}\label{fig_clip_demolition}
\end{figure}

Starting in 2017, the Moscow government has proposed to demolish many of the buildings constructed during the 60s and 70s and replace them with new housing blocks.\footnote{See \citet{Gunko2018} for details.} Figure~\ref{fig_clip_demolition} shows that the CLIP Q-scores for properties in our data that are scheduled for eventual demolition are clearly lower than the rest.

\begin{figure}
  \centering
  \includegraphics[scale=0.8]{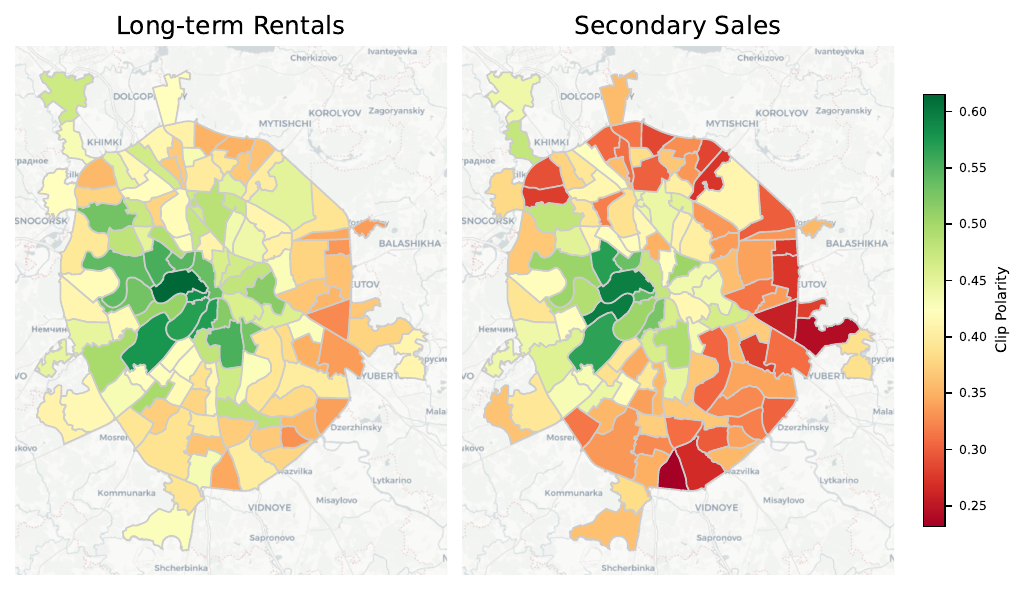}
  \caption{Moscow Districts and CLIP}\label{fig_clip_choropleth}
\end{figure}

Finally, in figure~\ref{fig_clip_choropleth} we show how CLIP Q-scores vary across Moscow districts. The city center and historically prestigious districts in the West have very high average scores. In contrast, relatively underdeveloped areas in the East and South of the city are, on average, in bad state of disrepair.

\section{Hedonic Model}

In this section, we present evidence that the CLIP Q-score measures characteristics of real estate units that are valued by the market. Specifically, we show that in a predictive model of property prices the CLIP Q-score is one of the most relevant features.

There is a long tradition of empirical investigations linking real estate values to observable characteristics.\footnote{\citet{Rosen_1974} defined \emph{hedonic prices} as the implicit prices of attributes of differentiated products.} A common approach is to specify a log-linear model of the form:

\begin{align*}
  \log p_i &= \alpha + \beta\cdot \mathrm{CLIP}_i + \boldsymbol{x}_i^\prime \boldsymbol{\delta} + \varepsilon_i
\end{align*}
where $p_i$ is either the asset price or a rental value for property $i$.\footnote{With normally working financial markets, asset prices and rental values are linked via the user cost of capital equation. Therefore, it is perfectly fine to estimate hedonic models using either.} Apart from the average adjusted CLIP Q-score for the property images, we also include a large number of other observable characteristics ($\boldsymbol x_i$).

In order to evaluate the predictive performance of CLIP, we randomly split the data into a ``training'' set (80\%) and a ``test'' set (20\%).\footnote{The random split is stratified by Moscow district.} We use the training data for estimation and the test data for evaluation.

\begin{figure}[htb!]
  \centering
  \includegraphics[scale=0.7]{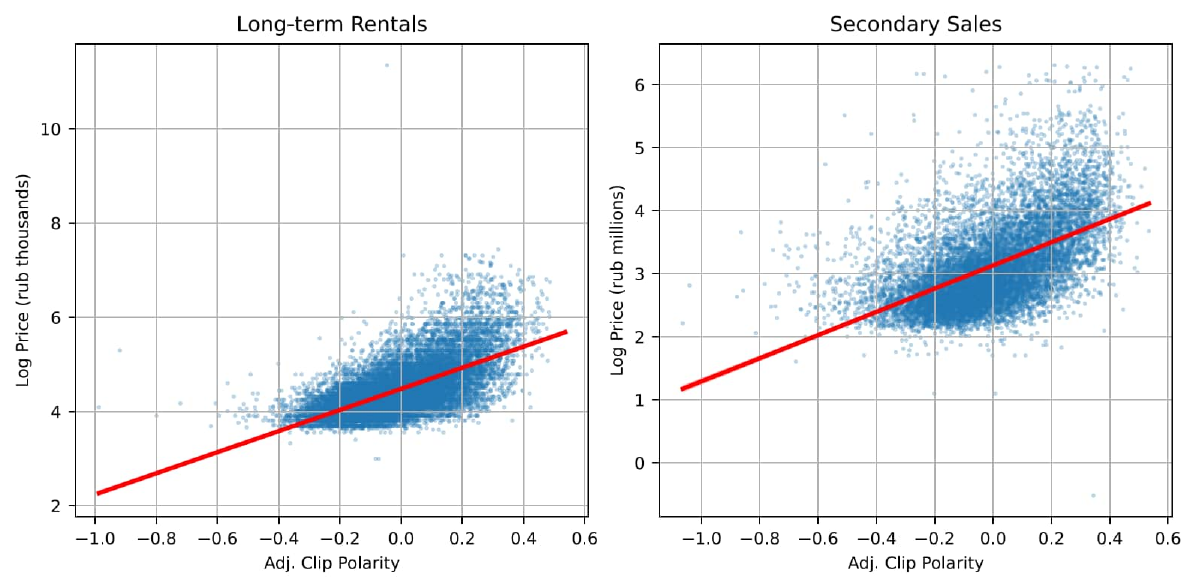}
      \vspace{-0.275in}
    {\tiny \flushleft \hspace{-1.2in} Notes: The price is the last quoted rental/sale asking price before the ad is closed on the platform.}
  \caption{CLIP and Prices}\label{fig_scatter_price_clip}
\end{figure}

Figure~\ref{fig_scatter_price_clip} shows the relationship between the adjusted CLIP Q-score and the last quoted price in the platform before the ad is closed.\footnote{Note that, if effect, what we have is the asking price and not the transaction price. Because it is reasonable to assume that better-looking properties give their owners more bargaining power when negotiating the final price, the findings in this section are probably conservative. In other words, the actual relationship between the CLIP score and the transaction price is bound to be stronger than the relationship we observe with the asking price.} There is a clear positive relationship, suggesting once again that the CLIP score is picking up signals of quality from the image data.

\subsection{Regression Results}

Table~\ref{tabl_hedonic_reg} presents a summary of results for the hedonic model estimated in log-linear form via OLS. For comparison, we present results with and without the CLIP Q-score.

\begin{table}[hbt!]
  \centering
  \caption{Hedonic Model Regression Results}\label{tabl_hedonic_reg}
    {\tiny
  \include{tabl_eval_hedonic_lr_ss_v0}

  }
\end{table}

The coefficient for the adjusted average CLIP polarity of the property's images is strongly positive and highly statistically significant. All else equal, an apartment with an average adjusted score just 0.1 units higher would be expected to rent (sell) for roughly 5.8\% (3.5\%) more.\footnote{A 0.1 higher CLIP score roughly equates to a 10\% higher probability assigned by the model to the ```high quality'' text description relative to the ``low quality'' text description for the images. This intuitive interpretation is arguably an additional positive feature of the approach.}

Both the in-sample (training set) and out-of-sample (test set) performance of the linear regression prediction model improve when the CLIP score is included in the model. For example, the adjusted R-squared improves by approximately 1 p.p.

\subsection{Gradient Boosting}

The log-linear functional form is convenient but definitely not dictated by theory. In order to check the performance of CLIP within a more flexible model, we trained a histogram-based gradient boosting algorithm.

Gradient boosting is an ensemble technique that builds a strong predictive model by sequentially combining multiple ``weak'' learners (simple decision trees). Each new tree is specifically designed to correct the errors made by all previous trees. Key hyper-parameters like the number of trees, the learning rate, and the tree depth were chosen by an extensive randomized grid search.

\begin{figure}[hbt!]
        \centering
        \includegraphics[scale=0.7]{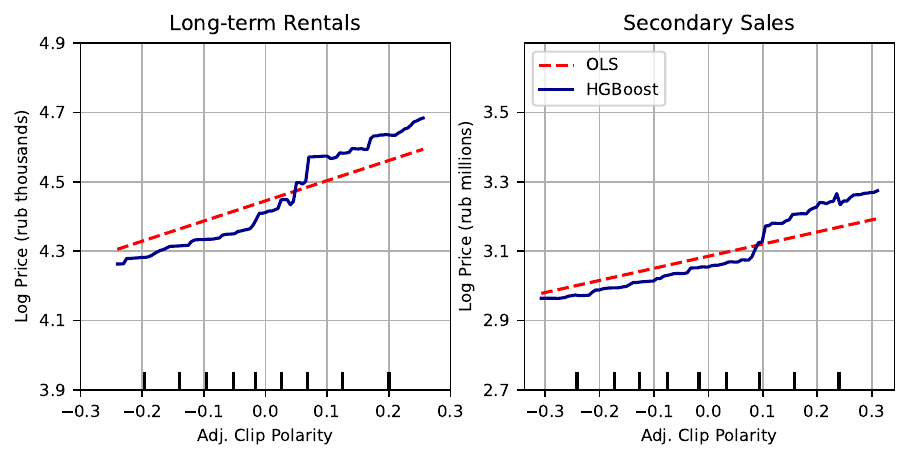}
    \caption{Partial Dependence Plots}\label{fig_pd_clip}
\end{figure}

Figure~\ref{fig_pd_clip} compares the conditional average predictions of the gradient boosting algorithm and the linear regression model on the test set. The simulations suggest the existence of mild non-linearities in the relationship between the CLIP Q-score and the expected log price.

\begin{table}[hbt!]
  \centering
  \caption{Gradient Boosting Model Results}\label{tabl_hedonic_histgradboost}
    {\tiny
  \include{tabl_eval_hedonic_hist_v0}

  }
\end{table}

Table~\ref{tabl_hedonic_histgradboost} provides the evaluation metrics for the gradient boosting model. Both the in-sample and out-of-sample performance are a bit better than the linear regression can attain. However, the relative improvements in performance when CLIP is included as a predictive feature is almost the same as with the linear model.

\subsection{Shapley Values}

Evaluating a model with and without a predictive feature provides only a partial picture regarding the importance of the feature in more general settings. To be concrete, in our application we have a very rich set of property descriptors other than the CLIP Q-score. It would be interesting to know whether CLIP is a desirable feature to have (or maybe invest in) in contexts with different data availability.

In the field of explainable machine learning, the widely accepted approach to determining the relative importance of different predictive features is the Shapley value decomposition \citep{Shapley_1953}.\footnote{Shapley proved that in the context of cooperative game theory, there is a single metric to determine the relative contribution of multiple players to a jointly beneficial task while respecting a number of intuitive axioms (efficiency, symmetry, additivity and null player).} In a nutshell, the Shapley value in the context of a prediction problem is a weighted average of the marginal contribution of a feature toward the improvement of a performance metric.

Formally, let $N$ be the set of all feature indexes: $N=\left\{1, 2, \dots, n \right\}$. Then the Shapley value for feature $j$ is given by:

\begin{align*}
\varphi_j =& \frac{1}{n}\sum_{S\subseteq N\backslash\{j\}} \binom{n-1}{|S|}^{-1} \left[v\left(S\cup \{j\}\right)-v\left(S\right)\right] \\
\end{align*}
where $v(S)$ is the performance attained with the model when the subset $S$ of features is included.\footnote{Intuitively, the Shapley value generalizes the approach in the previous section, where we compared the performance of the model with or without the CLIP feature while keeping all other predictors in. The calculation of $\varphi$ involve all possible combinations of the other features.}

\begin{center}
\vbox{
    \captionsetup{type=figure}
    \caption{Shapley Values}\label{fig_Shapley}
    \begin{subfigure}{0.98\textwidth}
        \centering
        \includegraphics[scale=0.7]{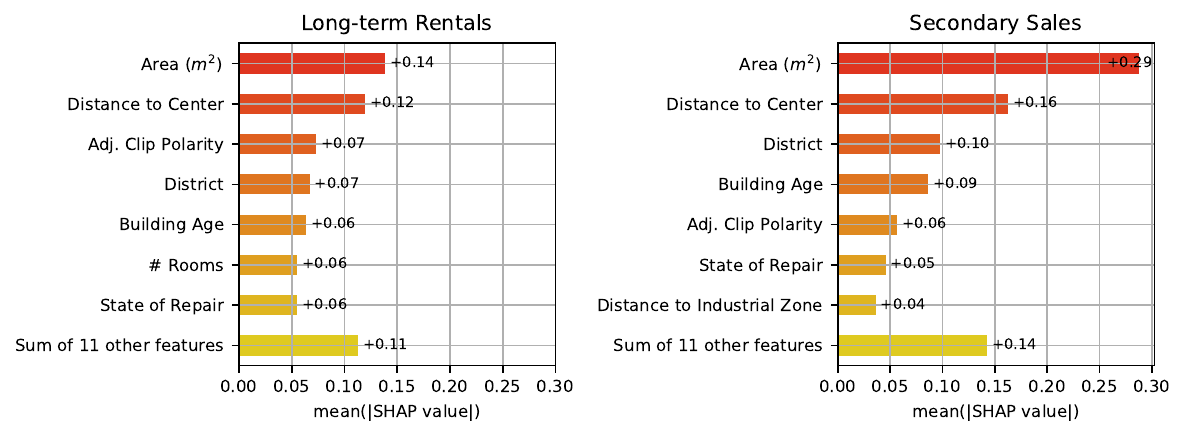}
        \caption{Linear Regression}
    \end{subfigure}
    \begin{subfigure}{0.98\textwidth}
        \centering
        \includegraphics[scale=0.7]{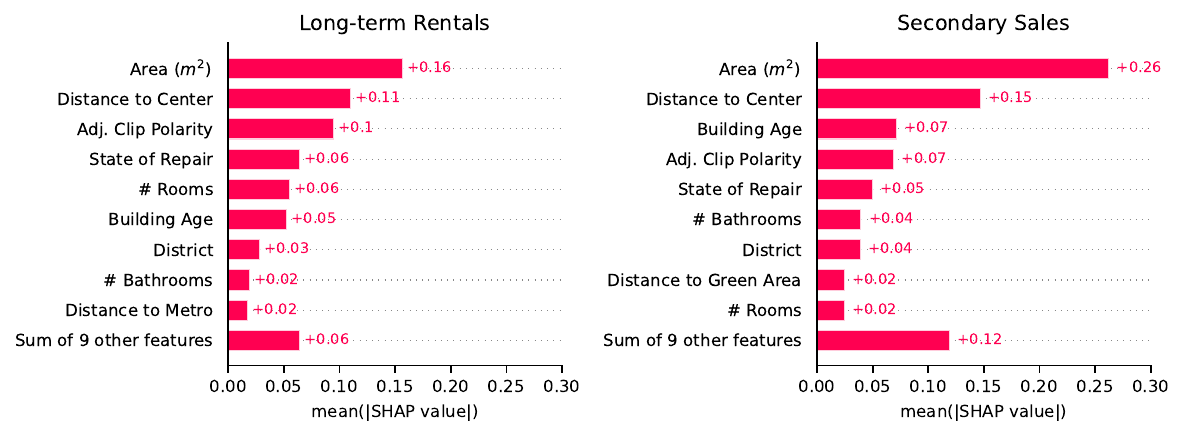}
        \caption{Gradient Boosting}
    \end{subfigure}
}
\end{center}

Figure~\ref{fig_Shapley} presents the results when the decomposition is applied to the linear regression (a) and the gradient boosting (b) models (the target metric is the negative RMSE). The property's total area and its distance to the city center are the two most important predictive features of its price. For rentals, the CLIP Q-score is in third place, while for sales it ranks fourth or fifth depending on the model.\footnote{Permutation importances ---an alternative way to assess predictive importance--- yield very similar results.}

In summary, the analysis in this section clearly shows that the CLIP Q-score captures important information regarding the properties that is not contained in other variables in the data.

\section{Time on the Market}

One important feature of our data gathering process is that we dedicated considerable effort to obtaining as accurate a picture as possible of the time the ad was on the platform before it was taken down. First, in order to avoid survivorship bias, we only included in our sample ads that are originally created during our data gathering period. Second, we re-visited ads periodically until we observed them closed. Finally, we kept re-visiting closed ads for an extra 14 days, in case they were re-opened.\footnote{Despite our best efforts, our limited infrastructure meant that a small fraction of ads were lost to follow-up without us being able to ascertain when they were closed. For this reason, the number of observations is slightly smaller in this section.}

\begin{table}[hbt!]
  \centering
  \caption{Time on the Market}\label{tabl_timeonmarket_descript}
    {\tiny

\include{tabl_timeonmarket_descriptives_v0}

  }
\end{table}

Table~\ref{tabl_timeonmarket_descript} presents descriptive statistics on observed durations.\footnote{We take license to refer to observed durations as `time on the market'. In reality, it is possible for the ad to be taken down while the property is still available for rent/sale.} Intuitively, sales ads are on the platform significantly longer than rentals. Consistent with this observation, we face much higher right-censoring of durations for the former (18.1\%) than for the latter (3.8\%).

\begin{figure}
    \centering
    \includegraphics[scale=0.7]{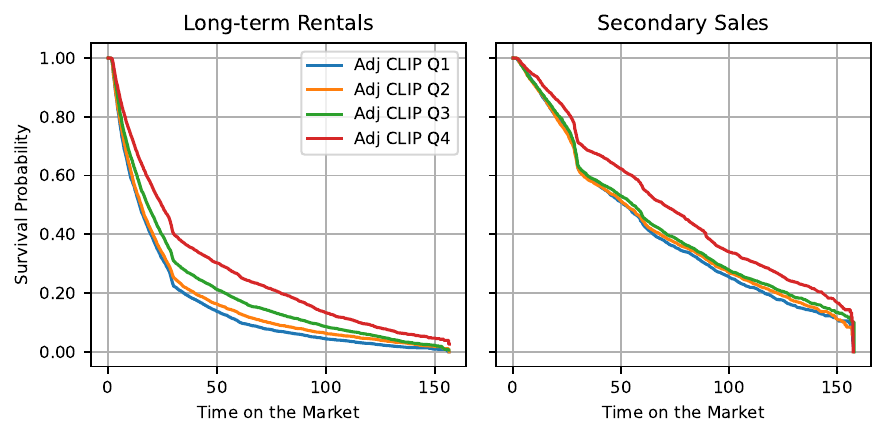}
    \caption{Kaplan-Meier Curves by CLIP Polarity Quartile}\label{fig_KM_by_clip}
\end{figure}

Figure~\ref{fig_KM_by_clip} presents Kaplan-Meier right-censoring-adjusted survival curve estimates for properties at different quartiles of the adjusted CLIP Q-score. The curves suggest that better looking properties take longer to rent/sell.

The reason for this somewhat puzzling result is given by the fact that better-looking properties are priced higher (as we learned in the previous section) and the higher-end of the real estate market moves slower. In order to show this point, we estimated a proportional hazard model in which time on the market is linked to both the CLIP Q-score and the initial asking price for the property:

\begin{align*}
  h(t_i\mid CLIP, p, \boldsymbol x) &= h_0(t_i) \exp\left(CLIP_i\cdot \gamma_1 \right)\exp\left(\log p^{init}_i \cdot \gamma_2 \right) \exp\left(\boldsymbol x_i^\prime \boldsymbol \theta \right)
\end{align*}
where $t_i$ is the observed duration for property $i$ and $h(\cdot)$ is the hazard function. For completeness, we include in the conditioning set the same full set of property descriptors ($\boldsymbol x_i$) as in previous analysis.\footnote{The results are qualitatively the same as long as the initial asking price is in the conditioning set.}

\begin{center}
\vbox{
    \captionsetup{type=figure}
    \caption{Proportional Hazard Model Results}\label{fig_PH_results}
    \begin{subfigure}{0.98\textwidth}
        \centering
        \includegraphics[scale=0.7]{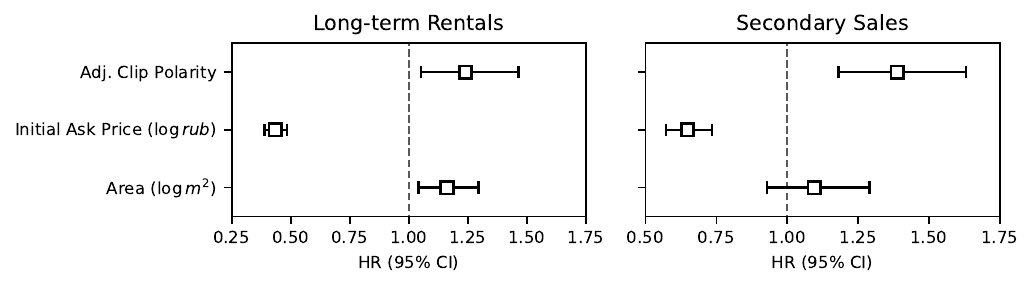}
        \caption{Hazard Ratios for Selected Features}
    \end{subfigure}
    \begin{subfigure}{0.98\textwidth}
        \centering
        \includegraphics[scale=0.7]{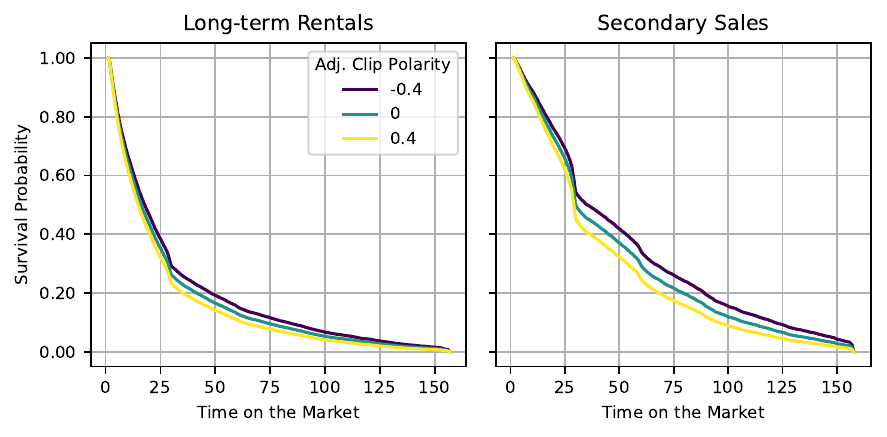}
        \caption{CLIP Average Partial Effects}
    \end{subfigure}
}
\end{center}

Figure~\ref{fig_PH_results} present the proportional hazard model results for the key features. The top panel (a) shows that the estimated hazard ratio for the adjusted CLIP polarity score is greater than one, implying that properties with higher scores sell \emph{faster} keeping the initial asking price and other characteristics fixed. The hazard ratio for the initial asking price is strictly below one, confirming our hypothesis.

The lower panel uses the model to simulate predicted survival curves keeping the adjusted CLIP Q-score at three meaningful levels (very low, medium, and high). In practical terms, the CLIP Q-score has an economically significant effect on sales, but not on rentals.

\section{Conclusions}

This paper introduced the CLIP Q-score, a simple, fast, and cost-effective approach to extracting product quality information from images. By leveraging an ensemble of pre-trained text and image deep neural networks, the methodology requires only a purposefully devised set of text descriptors from the researcher, bypassing complex training phases. Moreover, the resulting score is easy to interpret and fully reproducible.

We illustrated the approach through an extensive application using original real estate data from a major online platform. The resulting scores align with intuitive property attributes---such as neighborhood quality and recency of construction---and correlate tightly with AI model assessments.

By including the CLIP Q-score in a hedonic model, we show that it carries significant valuation-relevant information. The results are robust to functional form choice and hold in both rental and sales markets. We also show that the CLIP Q-score affects time on the market, although this relationship appears less economically consequential than the effect on prices.

We believe the insights from this and follow-up studies will offer new dimensions for search-theoretic and other economic models of consumer choice under imperfect information.

While our application focuses on a specific real estate market and timeframe, the underlying framework is highly adaptable and scales with ease. It is straightforward to envision this methodology generalizing well beyond property markets to other sectors where visual presentation influences buyer decisions. Consequently, this approach opens promising new avenues for future research across various branches of applied industrial organization and marketing science, offering a powerful tool for researchers analyzing digital marketplaces.

\newpage
\bibliography{bibdata}

\newpage\clearpage
\appendix\label{dataappendix}
\section{Appendix}
\clearpage
\setcounter{table}{0}
\renewcommand\thetable{\Alph{section}.\arabic{table}}
\setcounter{figure}{0}
\renewcommand\thefigure{\Alph{section}.\arabic{figure}}

\begin{table}
  \centering
    \caption{Querying LLaMA4 for Image Rating}\label{table_LLM_query}
      \begin{systembox}
        You are a helpful assistant that can look at pictures of real estate and provide accurate descriptions and analysis. Answer the user's questions regarding the attached images as clearly and accurately as possible.
    \end{systembox}    \vspace{-0.1cm}
    \begin{humanbox}
        Rate how luxurious and upscale this apartment looks based on the pictures. Give a score from 0 to 10 to each photo. Answer with a list of scores (e.g.: `$[4,3,5,7,2,0,9]$'). Omit any explanation or additional information.\\
        \newline
        \vspace{0.3cm}
        {\scriptsize
        \begin{tabular}{|c|c|c|}
        \toprule
         \texttt{Photo 1 URL} & \texttt{................} & \texttt{Photo 10 URL} \\
        \bottomrule
        \end{tabular}
        }
    \end{humanbox}

\end{table}

\begin{figure}
  \centering
  \includegraphics[scale=0.5]{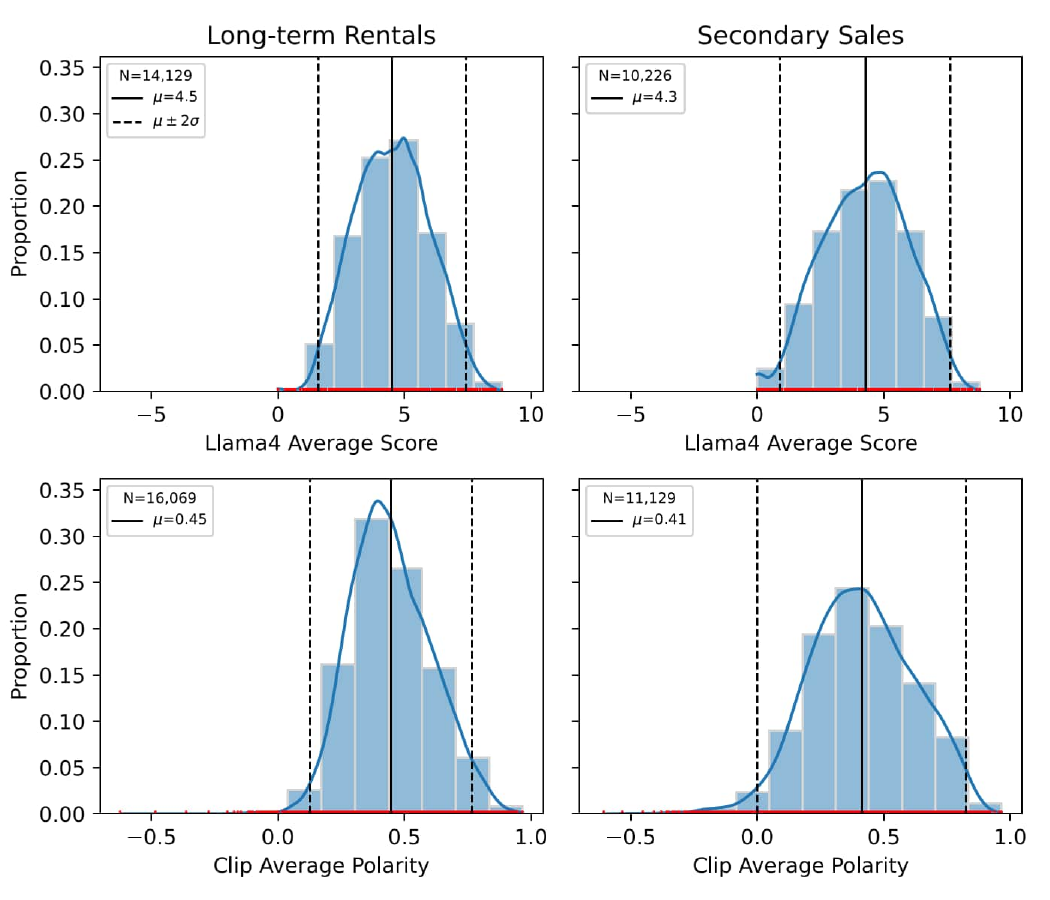}
  \vspace{-0.275in}
  {\tiny \flushleft \hspace{-0.9in} Note: Based on $N=$~16,047/11,129 ads with at least one photo.}
  \caption{Kernel Density Estimates --- LLaMA4 and CLIP Polarity Scores}\label{fig_LLama4_CLIP_kdensities}
\end{figure}

\end{document}

%% file: tabl_descriptives.tex
\begin{tabular}{lp{0.1in}cc}
\toprule
Variable & & Long-term Rentals & Secondary Sales \\
\midrule
\textbf{Area} ($m^2$) & & 54.859 & 61.845 \\
\textbf{Building Age} (years) & & 32.636 & 29.612 \\
\multicolumn{4}{l}{\textbf{\# Rooms}} \\
 \ \ \ \ 1  && 0.381 & 0.286 \\
 \ \ \ \ 2  && 0.404 & 0.380 \\
 \ \ \ \ 3  && 0.181 & 0.261 \\
 \ \ \ \ 4+ && 0.034 & 0.072 \\
\multicolumn{4}{l}{\textbf{\# Bathrooms}} \\
 \ \ \ \ 1  && 0.871 & 0.788 \\
 \ \ \ \ 2+ && 0.129 & 0.212 \\
\multicolumn{4}{l}{\textbf{Ceiling Height} ($m$)} \\
 \ \ \ \ 2.4--2.7 && 0.179 & 0.239 \\
 \ \ \ \ 2.7--3.0 && 0.155 & 0.193 \\
 \ \ \ \ 3.0+     && 0.107 & 0.261 \\
 \ \ \ \ NR       && 0.560 & 0.307 \\
\multicolumn{4}{l}{\textbf{'Apartment'}} \\
 \ \ \ \ Yes && 0.039 & 0.065 \\
\multicolumn{4}{l}{\textbf{Demolition Program}} \\
 \ \ \ \ Yes && 0.052 & 0.023 \\
\multicolumn{4}{l}{\textbf{Floor Number}} \\
 \ \ \ \ 1st  && 0.050 & 0.055 \\
 \ \ \ \ 2--9 && 0.606 & 0.578 \\
 \ \ \ \ 10--19 && 0.265 & 0.265 \\
 \ \ \ \ 20--29 && 0.057 & 0.067 \\
 \ \ \ \ 30+  && 0.021 & 0.034 \\
\multicolumn{4}{l}{\textbf{Construction Material}} \\
\ \ \ \ Brick && 0.165 & 0.162 \\
\ \ \ \ Blocks && 0.081 & 0.072 \\
\ \ \ \ Panels && 0.331 & 0.306 \\
\ \ \ \ Concrete && 0.226 & 0.354 \\
\ \ \ \ NR && 0.198 & 0.107 \\
\multicolumn{4}{l}{\textbf{State of Repair}} \\
\ \ \ \ No Repair && 0.011 & 0.188 \\
\ \ \ \ Cosmetic && 0.301 & 0.300 \\
\ \ \ \ European Style && 0.381 & 0.211 \\
\ \ \ \ Designer && 0.225 & 0.229 \\
\ \ \ \ NR && 0.081 & 0.072 \\
\multicolumn{4}{l}{\textbf{Geographic Features}} \\
\ \ \ \ Distance to Center && 111.054 & 109.099 \\
\ \ \ \ Distance to Metro && 7.033 & 7.213 \\
\ \ \ \ Distance to Green Area && 7.086 & 6.731 \\
\ \ \ \ Distance to Industrial Zone && 5.604  & 5.415 \\ \midrule
 \textbf{\# Ads} (images present)&& 16,047 & 11,129 \\
\bottomrule
\multicolumn{4}{p{3.4in}}{Notes: NR stands for ``not reported by the author''. Rooms include bedrooms and living/dining rooms. Distances are measured in 100m units.}\\
\end{tabular}

%% file: tabl_photo_descriptives.tex
\begin{tabular}{lcccccp{0.01in}ccccc}
\toprule
 & \multicolumn{5}{c}{Long-term Rentals} && \multicolumn{5}{c}{Secondary Sales} \\
 & mean & std. dev. & min & median & max && mean & std. dev. & min & median & max \\
\midrule
Aspect Ratio            & 1.034 & 0.385 & 0.345 & 0.750 & 5.000 && 1.107 & 0.391 & 0.291 & 1.332 & 5.224 \\
Resolution / $10^6$     & 0.822 & 0.254 & 0.043 & 0.691 & 1.638 && 0.865 & 0.257 & 0.048 & 0.737 & 1.637 \\
Sharpness / $10^3$      & 0.812 & 1.219 & 0.001 & 0.410 & 10.000 && 1.002 & 1.402 & 0.001 & 0.490 & 10.000 \\
Brightness / $10^2$     & 0.519 & 0.106 & 0.027 & 0.508 & 0.990 && 0.523 & 0.116 & 0.031 & 0.509 & 0.990 \\
BRISQUE / $10^2$        & 0.274 & 0.155 & 0.000 & 0.252 & 1.000 && 0.276 & 0.160 & 0.000 & 0.248 & 1.000 \\
CLIP Polarity           & 0.459 & 0.257 & -0.941& 0.460 & 0.995 && 0.425 & 0.309 & -0.961 & 0.435 & 0.994 \\
\bottomrule
\multicolumn{12}{p{5.4in}}{Notes: $N$ is 275,477 and 216,271 for Long-term Rentals and Secondary Sales resp. Aspect ratio is $width/height$. Resolution is the total number of pixels ($width\times height$). Sharpness is the variance of the Laplace transform of the grayscaled image (top-coded at 10K). Brightness is the mean of the lightness HSL channel.}\\
\end{tabular}

%% file: tabl_regression_clip_quality.tex
\begin{tabular}{lc@{\extracolsep{5pt}}c}
\toprule
  & \multicolumn{1}{c}{Long-term Rentals} & \multicolumn{1}{c}{Secondary Sales}  \\
  & (1) & (2) \\ \midrule
 Aspect Ratio & 0.074$^{***}$ & 0.119$^{***}$ \\
& (0.004) & (0.006) \\
 Resolution / $10^6$ & -0.015$^{**}$ & -0.053$^{***}$ \\
& (0.006) & (0.008) \\
 Brightness / $10^2$ & 0.236$^{***}$ & 0.286$^{***}$ \\
& (0.009) & (0.013) \\
 BRISQUE / $10^2$ & -0.123$^{***}$ & -0.056$^{***}$ \\
& (0.005) & (0.008) \\
 Sharpness / $10^3$ & -0.014$^{***}$ & -0.009$^{***}$ \\
& (0.001) & (0.001) \\ \midrule
 Observations       & 275,477 & 216,271 \\
 $R^2$              & 0.025   & 0.030 \\
 F-Stat & 321.2$^{***}$   & 277.5$^{***}$  \\
\bottomrule
\multicolumn{3}{p{3in}}{\textit{Notes:} Standard errors clustered at the property ad level in parenthesis. $^{***}$p$<$0.01. Constant term omitted.} \\
\end{tabular}

%% file: tabl_eval_hedonic_lr_ss_v0.tex
% This is based on UNADJUSTED CLIP. IT COMBINES LR AND SS EVAL TABLES

\begin{tabular}{l@{\extracolsep{35pt}}cc@{\extracolsep{35pt}}cc}
  \toprule
         & \multicolumn{2}{c}{Long-term Rentals} & \multicolumn{2}{c}{Secondary Sales} \\
         &  (1) & (2) &  (3) & (4)  \\ \midrule
    CLIP & \xmrk & \chkmrk & \xmrk & \chkmrk  \\
  \midrule
  \multicolumn{5}{l}{\emph{Coefficient Estimate}} \\[1pt]
  Avg. Adj. CLIP polarity   &  & 0.582$^{***}$ & & 0.350$^{***}$ \\
  (HC3 S.E.)           &  & (0.018)       & &(0.019)        \\[3pt]
  \midrule
  \multicolumn{5}{l}{\emph{Train-sample Performance}} \\[1pt]
  Adj-R2 & 0.869 & 0.885 & 0.902 & 0.908 \\
  RMSE   & 0.181 & 0.170 & 0.195 & 0.189 \\[3pt]
  \multicolumn{5}{l}{\emph{Test-sample Performance}} \\[1pt]
  Adj-R2 & 0.874 & 0.886 & 0.897 & 0.904 \\
  RMSE   & 0.179 & 0.170 & 0.192 & 0.186 \\[3pt]
  \multicolumn{5}{l}{\emph{Other Included Features}} \\
  Unit        & \chkmrk & \chkmrk & \chkmrk & \chkmrk \\
  Building    & \chkmrk & \chkmrk & \chkmrk & \chkmrk \\
  Geography   & \chkmrk & \chkmrk & \chkmrk & \chkmrk \\
  District FE & \chkmrk & \chkmrk & \chkmrk & \chkmrk \\
  \bottomrule
  \multicolumn{5}{p{4.9in}}{Notes: train sets N=12,725/8,812; test sets N=3,182/2,204. Unit level features: area (quadratic), room number, bathroom number, ceiling height, floor number, repair type, number of images. Building level: age (quadratic), material, demolition. Geography: distance to center/metro/green area/industrial zone.}\\
\end{tabular}

%% file: tabl_eval_hedonic_hist_v0.tex
\begin{tabular}{l@{\extracolsep{35pt}}cc@{\extracolsep{35pt}}cc}
  \toprule
         & \multicolumn{2}{c}{Long-term Rentals} & \multicolumn{2}{c}{Secondary Sales} \\[1pt]
                  &  (1) & (2) &  (3) & (4)  \\ \midrule
    CLIP & \xmrk & \chkmrk & \xmrk & \chkmrk \\[3pt]
  \midrule
  \multicolumn{5}{l}{\emph{Train-sample Performance}} \\[1pt]
    Adj-R2 & 0.910 & 0.933 & 0.949 & 0.959 \\
    RMSE   & 0.151 & 0.130 & 0.143 & 0.127 \\[3pt]
  \multicolumn{5}{l}{\emph{Test-sample Performance}} \\[1pt]
    Adj-R2 & 0.887 & 0.899 & 0.915 & 0.923 \\
    RMSE   & 0.173 & 0.164 & 0.181 & 0.172 \\
  \bottomrule
  \multicolumn{5}{p{4.0in}}{Notes: Unit, building, geographic, and district level features included.}\\
\end{tabular}

%% file: tabl_timeonmarket_descriptives_v0.tex
\begin{tabular}{lcccccc}
  \toprule
  Status  & Count & Mean  & Std.Dev. & Min & Median & Max \\[1pt]
  \midrule
  \multicolumn{7}{l}{\emph{Long-term Rentals}} \\[2pt]
  Closed   & 15,208 & 27.5 & 30.0 & 1.2 & 16.4  & 156.6   \\
  Censored & 603  & 126.5 & 17.7 & 60.5 & 126.4 & 156.6   \\[1pt]
  \multicolumn{7}{l}{\emph{Secondary Sales}} \\[2pt]
  Closed   & 8,981 & 51.3  & 37.1 & 1.3  & 40.6  & 157.8 \\
  Censored & 1,990 & 128.7 & 16.7 & 60.0 & 128.5 & 157.4 \\
  \bottomrule
\end{tabular}